\documentclass[aps,floatfix,prc,showpacs,showkeys,amsmath,amssymb,superscriptaddress, twocolumn]{revtex4-1}
\usepackage{graphicx}
\usepackage{xcolor}
\usepackage{ulem}
\usepackage{hyperref}
\usepackage{amssymb}

\renewcommand\sout{\bgroup \color{blue}\ULdepth=-.5ex \ULset}
\begin{document}

\begin{flushright}
CERN-TH-2020-141
\end{flushright}
\vskip 0.3cm
\title{Multiplicity dependence of (multi-)strange baryons in the canonical ensemble with
phase shift corrections}
\author{Jean Cleymans}
\affiliation{UCT-CERN Research Centre and Department of Physics, University of Cape Town, Rondebosch 7701, South Africa}
\author{Pok Man Lo}
\affiliation{Institute of Theoretical Physics, University of Wroclaw, Wroclaw, Poland}
\author{Krzysztof Redlich}
\affiliation{Institute of Theoretical Physics, University of Wroclaw, Wroclaw, Poland}
\affiliation{Theoretical Physics Department, CERN, CH-1211 Gen\`eve 23, Switzerland}
\author{Natasha Sharma}
\affiliation{Department of Physics, Panjab University, Chandigarh 160014, India}
\date{\today}
\begin{abstract}
The increase in strangeness production  with charged particle
multiplicity, as seen by the ALICE collaboration at CERN 
in p-p, p-Pb and Pb-Pb collisions, is
investigated in the hadron resonance gas model taking into account
interactions among hadrons using S-matrix
corrections  based on known phase shift analyses.
Strangeness  conservation  is taken into account in the framework of
the canonical strangeness ensemble.
A very good description is obtained for the variation of the strangeness content in
the final state as a function of the
number of charged hadrons in the mid-rapidity region  using the same 
fixed temperature value as  obtained in the most central Pb-Pb collisions and with a fixed strangeness suppression factor $\gamma_s = 1$. 
It is shown that the number of charged hadrons 
is linearly proportional to the volume of the system.  For small multiplicities  the canonical ensemble  with local strangeness conservation restricted to mid-rapidity
leads to  a stronger
suppression of (multi-)strange baryons than seen in the data. 
This is compensated by introducing a global conservation of strangeness in the whole phase-space which is parameterized  by the  canonical correlation volume  larger than the  fireball volume at the mid-rapidity. 
The results on comparing the hadron resonance gas model with and without
S-matrix corrections, are presented in detail.
It is shown that the interactions introduced by the phase shift
analysis via the S-matrix formalism are essential for a better  
description of the yields data. 
\end{abstract}
\maketitle

%
\section{\label{secIntroduction}Introduction}

The analysis of data on hadron  yields produced in heavy-ion collisions covering a broad range of 
energies in fixed-target and collider experiments,   confirms  that produced hadrons originate from a thermal fireball formed 
in such collisions~\cite{Andronic:2017pug,Stachel:2013zma,Andronic:2014zha,Becattini:2010sk,Andronic:2018qqt,Chatterjee:2015fua,Das:2016muc,
Abelev:2013haa}.
The yields of produced hadrons are quantified by the statistical operator of the hadron resonance gas model (HRG) with a common freezeout temperature $T_f$ and chemical potentials $\vec\mu_f$  associated with the conserved charges. The volume of the fireball is fixed such as to reproduce the multiplicities of  hadrons at a given collision energy $\sqrt s$. The thermal freezeout parameters of the produced fireball were shown to be uniquely linked to the collision energy  \cite{Andronic:2017pug,Cleymans:1999st}.

The description of particle production in nucleus-nucleus collisions in the framework of the HRG is particularly transparent 
 at the Large Hadron Collider (LHC) 
energies. There, at midrapidity, particles and antiparticles are produced in pairs, thus all chemical potentials vanish. Consequently, the chemical freezeout of all hadrons is quantified by the temperature and the volume of the fireball only.

An impressive overall agreement has been obtained between the measured particle yields by the ALICE collaboration for the most central Pb–Pb collisions and the HRG model results~\cite{Andronic:2017pug,Stachel:2013zma}. The agreement spans nine orders of magnitude in abundance values, encompassing all measured mesons and baryons, as well as light nuclei and hypernuclei and their antiparticles. The analysis was further successfully extended to heavy flavor production by accounting for the initially produced  charm quark pairs and  their conservation laws~\cite{Andronic:2017pug,Andronic:2019wva,Andronic:2018vqh}.

For the most central Pb–Pb collisions, the best description of the ALICE data on yields of particles in one unit of rapidity at mid-rapidity was obtained at $T_{f} = 156.6\pm   1.7$ MeV \cite{Andronic:2017pug,Andronic:2018qqt}. 
Remarkably, this  value of  $T_{f}$ coincides within errors with   the pseudo-critical temperature
$T_c = 156.5 \pm  1.5$  MeV obtained from first principles Lattice QCD  (LQCD) calculations \cite{Bazavov:2018mes}, albeit with the possibility of a 
broad transition region~\cite{Borsanyi:2020fev}.
Furthermore, assuming that the net-charge probability distributions follow the Skelam~\cite{BraunMunzinger:2011dn} or 
generalized Skelam distribution function \cite{BraunMunzinger:2011ta}, the ALICE data on 
different particle yields have been directly compared with  LQCD results on  charge fluctuations  and 
correlations \cite{Braun-Munzinger:2014lba}. 
The results of this comparison have shown  that the susceptibilities are consistent within errors with the results 
obtained from LQCD  at the chiral crossover point. 
This  provides strong evidence for the observation that in central Pb-Pb collisions all hadrons and their bound states are originating from the hadronized QGP. 
This conclusion can also be extended to lower collision  energies, 
since freezeout conditions in central heavy-ion collisions were shown to closely follow the 
chiral crossover  at finite baryon chemical potential as calculated in LQCD.
A comparison between LQCD and HRG has also been presented in~\cite{Becattini:2016xct} where a higher estimate
for the pseudo-critical temperature was obtained.

One of the consequences of confinement in QCD is that at lower temperatures $T\leq T_c$,   physical observables require  a representation in terms of hadronic states. On the other hand,  the successful description of hadron yields in heavy-ion collisions by the HRG and the coincidence of the  freezeout and chiral crossover temperatures suggests that the statistical operator of HRG is a good approximation to QCD thermodynamics in the confined phase. Indeed, a direct comparison of the equation of state (EOS)  of  LQCD and the HRG model has shown that they closely coincide in the hadronic phase, both at vanishing and at small finite $\mu_B$ \cite{Bazavov:2017dus,Noronha-Hostler:2019ayj}.
These results provide  strong support for the view that  matter produced in central heavy-ion collisions is a  QCD medium in thermal equilibrium
described by  the HRG statistical operator in the hadronic phase.   There are, however, some limitations to the HRG description of QCD thermodynamics which are identified in the context of  LQCD  as well as in the description of   hadron production yields data at the LHC in heavy-ion collisions \cite{Bazavov:2014xya,Lo:2017lym,Andronic:2018qqt}.

Recent results of LQCD on second-order fluctuations and correlations $\chi_{PQ}$ of conserved charges $P$ and $Q$  allow to identify the HRG approximation in different sectors of hadronic quantum numbers. 
In particular, in this context it was shown,  that at $T\simeq T_{c}$ the HRG underestimates the baryon-strange $\chi_{BS}$ and overestimates the baryon-charge $\chi_{BQ}$  correlations \cite{Lo:2017lym,Fernandez-Ramirez:2018vzu}.
In the HRG, the main contribution to $\chi_{BQ}$ is due to protons and resonances which decay into protons, thus too large  a value of $\chi_{BQ}$ indicates an excess of protons at $T\simeq T_{c}$. 
Indeed the HRG analysis of ALICE data in central heavy-ion collisions at collision energy $\sqrt s=2.72$ TeV has shown, that the HRG predicts about $25\%$  more protons and antiprotons than measured by the ALICE collaboration in central Pb–Pb collisions at the LHC \cite{Andronic:2017pug,Andronic:2018qqt}. 
This constitutes the much-debated “proton-yield anomaly” in the heavy-ion collisions at the LHC.

The partition function of the HRG  is evaluated in the grand canonical (GC) ensemble as a mixture of ideal gases of 
all stable hadrons and resonances as reported by the Particle Data Group (PDG)~\cite{Tanabashi:2018oca}. 
A very comprehensive analysis of the influence of hadronic resonances expected from lattice QCD but not listed in the PDG
has been presented in~\cite{Bazavov:2014xya,Alba:2020jir}.

In the spirit of the S-matrix formalism  \cite{Dashen:1969ep,Venugopalan:1992hy,Weinhold:1997ig,Lo:2017lym,Dash:2018mep,Dash:2018can}, which provides a theoretical framework to implement interactions in a dilute many-body system in equilibrium, the presence of resonances corresponds to attractive interactions among hadrons. 
However, many of the simplifying assumptions of attractive interactions implicit in the HRG model are not necessarily consistent with hadron scattering data.  
In particular, for an accurate determination of interaction effects, a proper  resonance invariant mass distribution and the presence of many non-resonant contributions have to be included to be consistent with scattering data.
This can be done systematically within the S-matrix approach where two-body interactions are, via the empirical scattering phase shifts, included 
to construct the leading interaction term in the virial expansion of the grand canonical 
potential \cite{Dashen:1969ep,Venugopalan:1992hy,Weinhold:1997ig,Giacosa:2016rjk,Lo:2017sde,Dash:2018can,Dash:2018mep,Friman:2015zua, Dash:2018mep}. The resulting interacting density of states is then folded into an integral over thermodynamic distribution functions,  which, in turn, yields the contributions from interactions to a particular thermodynamic quantity.

The S-matrix approach has been applied to study the baryon-charge susceptibility $\chi_{BQ}$ in a thermal medium \cite{Lo:2017lym}.  It was demonstrated that the implementation of the empirical pion-nucleon phase shifts is crucial for the proper interpretation of the LQCD result.  Also in the analysis of observables involving nucleons in ultra-relativistic nucleus-nucleus collisions, a careful treatment of pion-nucleon interactions in the partition function of the HRG model  could resolve the proton-yield anomaly \cite{Andronic:2018qqt}. 
Furthermore, in the strange baryon sector of HRG, the improvement of interactions within the
S-matrix formalism was shown to increase the strange-baryon correlations $\chi_{BS}$ towards the  LQCD value at $T\sim T_{c}$ \cite{Fernandez-Ramirez:2018vzu}. 
The coupled-channel analysis naturally incorporates some additional hyperon states beyond those listed in the PDG~\cite{Tanabashi:2018oca}, thus supporting the importance of these states in explaining the LQCD results~\cite{Bazavov:2014xya,Lo:2017ldt}

It is clear that the S-matrix scheme can improve the HRG model in approximating the QCD partition function in the hadronic phase, thus producing a more accurate description of the measured particle yields in heavy-ion collisions.

\begin{figure*}[ht]
\centering
\includegraphics[width=0.49\linewidth]{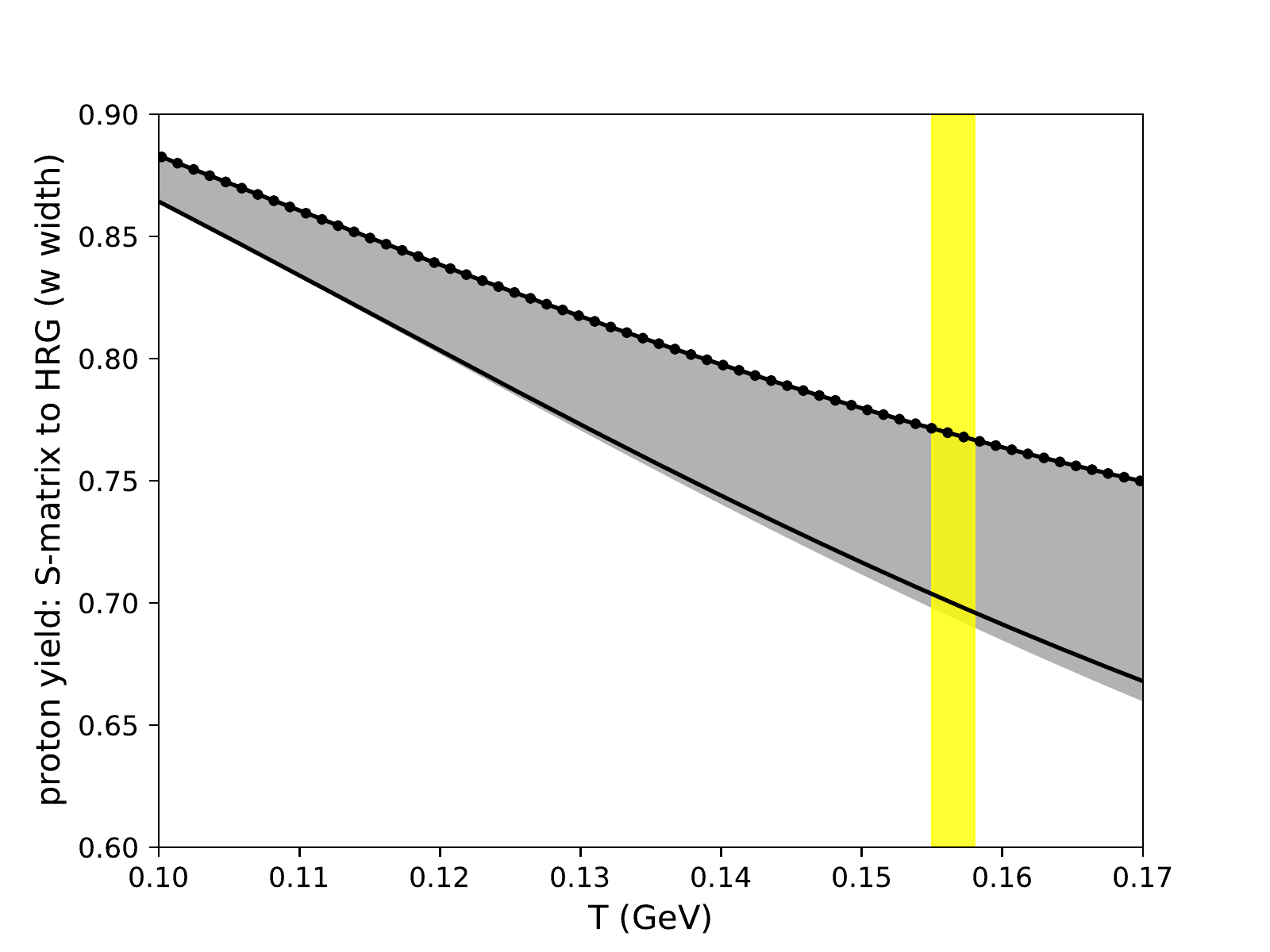}
\includegraphics[width=0.49\textwidth]{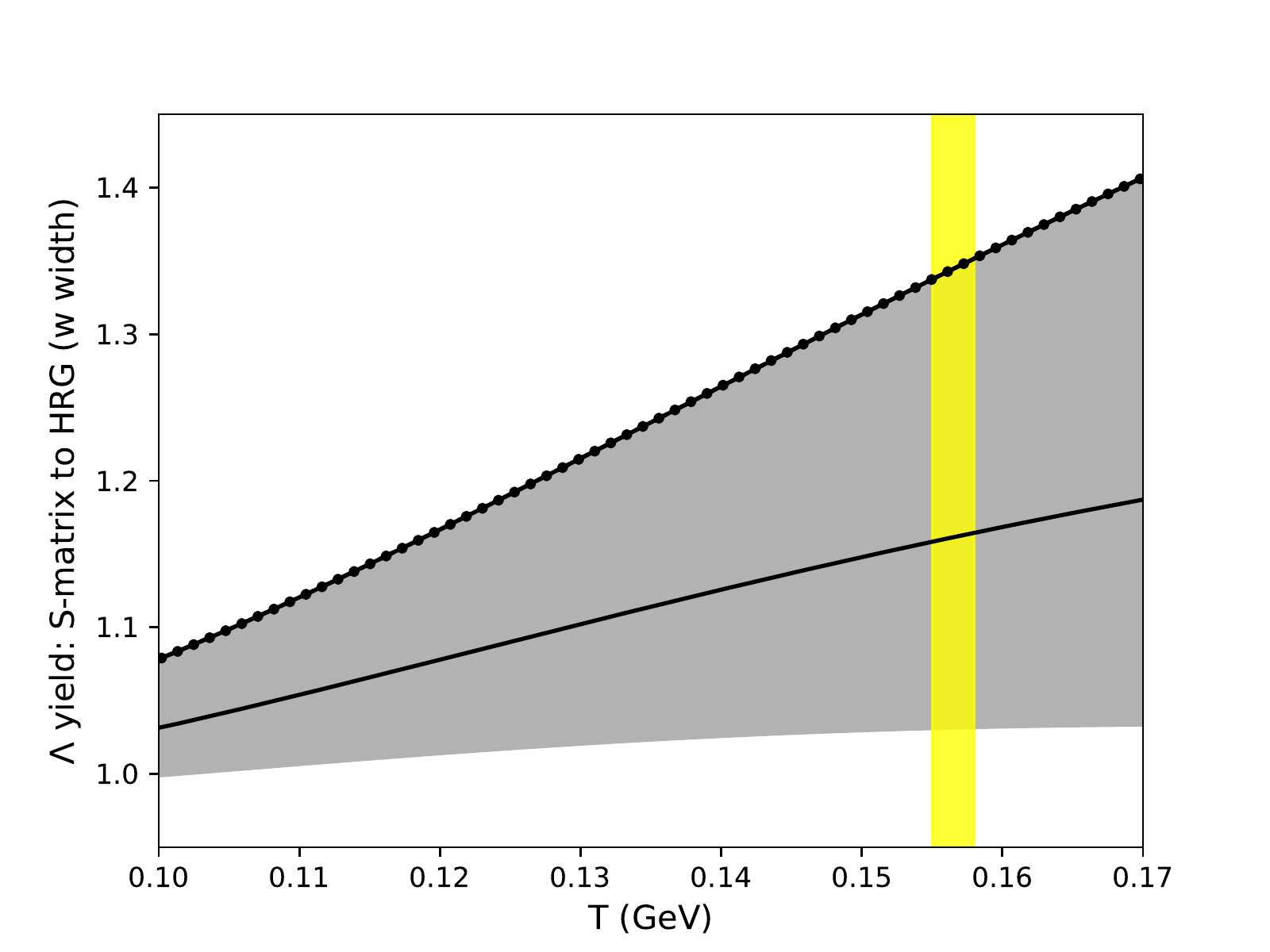}
	\caption{
The ratios of proton (left) and $\Lambda+\Sigma^0$ (right) yields in the S-matrix and the HRG baseline results.
The yellow band signifies the range of pseudo-critical temperatures $T_c = 156.5 \pm 1.5$ MeV \cite{Bazavov:2018mes}.
The shaded region (in gray) shows the S-matrix scheme implemented with different levels of approximation: e.g. from including only elastic and quasi-elastic scatterings (black solid line), to the full list of channels (black points). 
See Table~\ref{tab1} and the discussion in Sec.~\ref{sec3}.
These results can be interpreted as the correction factors for a statistical model provided that the same HRG baseline is employed. (Here it is THERMUS in its default setting ~\cite{Wheaton:2004qb}.)
In all figures we follow the convention to use $\Lambda$ to signify $\Lambda+\Sigma^0$ unless otherwise specified.
}
\label{fig:fig1}
\end{figure*}

A qualitative display of the S-matrix corrections (relative to the HRG baseline) is shown in Fig.~\ref{fig:fig1}. 
The trend of the overall correction is clear:
a reduction in the proton yield and an enhancement in the $\Lambda+\Sigma^0$ yields.
The corrections can reach $\approx -25 \%$ ( $\approx +23 \%$) for protons ($\Lambda+\Sigma^0$-baryons) at the LHC freezeout conditions.
The shaded region (in gray) shows the S-matrix scheme implemented with different levels of improvement: e.g. from including only elastic and quasi-elastic scatterings (black solid line), to the full list of channels (black points). 
A detailed composition analysis of the hadron yields under the S-matrix scheme will be presented in Sec.~\ref{sec3}.

In the usual formulation of the HRG model, the  conservation laws are implemented in the GC ensemble where they are fulfilled on 
average and are controlled via chemical potentials linked to the conserved quantum numbers. 
In heavy-ion collisions, however, the conservation of quantum numbers is fulfilled exactly as they are fixed by the initial conditions. 
Focusing on strangeness production at the LHC, the total strangeness $S$  is exactly zero in the full phase-space. 
Consequently,  the constraint of $S=0$, can in general  influence strangeness production  in the  acceptance region, which is usually a slice of one unit of rapidity, at mid-rapidity. 
The exact and global conservation of net-baryon  numbers was also recently shown to be crucial when discussing 
fluctuations of conserved charges 
in heavy-ion collisions 
in a given acceptance region~\cite{Braun-Munzinger:2020jbk,Bzdak:2012an,Braun-Munzinger:2018yru}.

To account for exact charge conservation the HRG  thermodynamic potential has to be formulated in the 
canonical (C) 
ensemble~\cite{Hagedorn:1971mc, Hagedorn:1984uy,Cleymans:1990mn,Cleymans:1998yb,Hamieh:2000tk,Ko:2000vp,BraunMunzinger:2003zd,Begun:2004gs} 
which implies  suppression of charged particle yields relative to their GC values. 
The HRG model formulated in C-ensemble has provided a very useful framework for the centrality  and 
system-size dependence  of particle production in this particularly strangeness production and 
suppression~\cite{Cleymans:1998yb,Hamieh:2000tk,Redlich:2001kb}. 
The applicability of the model in small systems like p-p collisions \cite{Kraus:2008fh}, p-A \cite{Sharma:2018owb}  
and $e^+e^-$ annihilation has also been successfully discussed in the 
literature \cite{Andronic:2008ev,Becattini:2008tx}.

Recently the ALICE collaboration has observed an interesting systematic behavior in  particle production yields  at LHC energies, pointing out that the ratios of  
identified particles to pions  depend solely on the  charged-particle multiplicity $dN_{ch}/d\eta$, regardless of system type and collision energy  \cite{Acharya:2020zji,ALICE:2017jyt,Abelev:2013haa,Adam:2015vsf,Abelev:2013vea,Abelev:2013xaa,Abelev:2013zaa}. It is particularly  interesting in these data that the evolution of  strange  particle yields  with $dN_{ch}/d\eta$ shows  patterns that are common  in p-p,   p-Pb and Pb–Pb collisions. The integrated yields of strange and multi-strange particles, relative to pions increases  with the charged-particle multiplicity and the enhancement  becomes more pronounced with increasing strangeness content. Such a pattern of enhancement of (multi-)strange hadrons with the number of charged particles is qualitatively similar to what has been   observed previously by the WA97 and NA57  collaboration at SPS energies at CERN \cite{Antinori:2001qk,Aggarwal:2010ig} and by the STAR collaboration at RHIC~\cite{Adamczyk:2017iwn,Cleymans:2011pe}.

The  strangeness enhancement in high-multiplicity p-p collisions observed by the ALICE collaboration and the relation of  these data to p-A and A-A collisions
 cannot be explained quantitatively  by general purpose QCD Monte Carlo models \cite{ALICE:2017jyt,Acharya:2019kyh}. 
 On the other hand, the first analysis of data within the thermal model with an exact strangeness conservation has shown that the ALICE data are  following the model expectations \cite{Vislavicius:2016rwi,Sharma:2018owb,Adam:2015vsf,Kalweit:2016lrl,Anielski:2014}.
 Actually, the observed  properties  of multi(strange) hadrons with charged-particle multiplicity was 
predicted in the context of  the thermal model as being due to the canonical suppression  
effect \cite{Hamieh:2000tk,Redlich:2001kb}.  
 Indeed, 
at the LHC, the canonical suppression  depends on the total density of strange particles $n_s(T)$ and the 
correlation volume parameter  $V_C$ which quantifies the range  of strangeness  conservation and can be  
parameterized by  $dN_{ch}/d\eta$. For large $V_C$,
the canonical suppression factor tends to unity, and particle yields are approaching their   GC values. This is e.g. the case for central heavy-ion collisions at the LHC, where the  GC HRG  provides an excellent description of the yields. However,  in events with small charged particle multiplicities where $V_C$ is small, the canonical corrections cannot be neglected. 
The characteristic prediction of the HRG model in the  C-ensemble was  an increasing suppression of strange  particle yields with decreasing collision energy and  collision centrality, as well as with  increasing strangeness  content of hadrons \cite{Hamieh:2000tk,Redlich:2001kb}.

The main objective of this paper is to apply the S-matrix extended HRG model to  analyze data obtained by the  ALICE 
collaboration on charged-particle multiplicity in p-p collisions at 7 TeV~\cite{ALICE:2017jyt} as well as
13 TeV \cite{Acharya:2019kyh},  
in  p-Pb collisions at 5.02 TeV~\cite{Abelev:2013haa,Adam:2015vsf} and in  Pb-Pb collisions
at 2.76 TeV~\cite{Abelev:2013vea,Abelev:2013xaa,Abelev:2013zaa},  all in  the central region of rapidity.  

We focus on yields of (multi-)strange hadrons and discuss strangeness suppression as a function of $dN_{ch}/d\eta$ and strangeness content of hadrons. To this end, we formulate the HRG model in the canonical ensemble of strangeness conservation and account for differences between the fireball volume  at midrapidity $V_A$  and the correlation volume $V_C$ required for exact global strangeness conservation
\cite{Hamieh:2000tk,Hagedorn:1984uy,Satz:2017ltg,Castorina:2013mba}. We include the S-matrix corrections to the hyperon yields,  employing an existing coupled-channel study involving  $\pi \Lambda$,  and $\pi \Sigma$  interactions 
in the $S = -1$ sector. We  include the S-matrix corrections to proton production by using the  empirical phase shifts of $\pi N$ scattering 
and the contribution of hyperons to proton yield.

 We will show that the yields of (multi-)strange baryons versus $dN_{ch}/d\eta$ measured by the ALICE collaboration follow the expectations of thermal production with the canonical suppression due to exact strangeness conservation  at  fixed temperature  $T\simeq 156.5$ MeV,  that is consistent with the chiral crossover  in LQCD and with a fixed strangeness suppression factor $\gamma_s = 1$. 
 The yields of (multi-)strange hadrons are quantified for different $dN_{ch}/d\eta$   within one standard deviation.  
 Furthermore, increasing suppression with increasing strangeness content of baryons as
 a function of $dN_{ch}/d\eta$  follows patterns  obtained recently by the ALICE collaboration. We will  discuss the S-matrix extended HRG model  results in the C-ensemble on yields of   (multi-)strange baryons, protons and kaons  normalized to pion multiplicity for different  $dN_{ch}/d\eta$,  as well as,
 normalized to  $\Lambda$ yield to indicate the importance of S-matrix corrections.
 Our results show, that the thermal origin of particle yields observed in the most central heavy-ion collisions at the LHC \cite{Andronic:2017pug} can be extended to events with decreasing $dN_{ch}/d\eta$. Furthermore, the observed scaling of hadron yields with $dN_{ch}/d\eta$ for different colliding systems is a natural consequence of the HRG model with exact conservation of   strangeness. These results provide further evidence for the thermal origin of particle production at the LHC in p-p, p-A and A-A collisions at a common $T_f\simeq T_{c}$.
 
The paper is organized as follows: In Section II,  we introduce the HRG model in the canonical ensemble. In Section III,  we discuss  the S-matrix corrections to proton and $\Lambda$ yields. In Section IV, we introduce the   model comparison with LHC data. In section V, we present our summary and conclusions.

\section{Strangeness production with canonical suppression}
In the  GC ensemble, the quantum numbers are conserved on  average and  are  implemented  using the corresponding  chemical potentials
$\vec\mu$ linked to conserved charges $\vec Q$.
The partition function depends on thermodynamic quantities and the Hamiltonian describing the system, 

\begin{equation}
\label{eq1}
Z_{GC} = \textrm{Tr} \left[ e^{-(H-\vec \mu\cdot \vec Q)/T}\right]. 
\end{equation}
 In the framework of the HRG model considered here and in the Boltzmann approximation,  the above  leads to
\begin{equation}
\label{lnZ}
\ln Z_{GC}(T,\vec \mu,V) = \sum_i g_i V \int\frac{d^3p}{(2\pi)^3}\exp\left( -\frac{E_i- {\vec q}_i{\vec\mu} }{T} \right)\\,
\end{equation}
where $g_i$ is the spin degeneracy factor of particle $i$ and $E_i$ its energy, $V$ is the fireball volume, ${\vec \mu}_i$
are  chemical potentials associated with  conserved charges  $\vec{q}_i$ carried by particle $i$. The sum is taken over all stable hadrons and resonances as well as their antiparticles. 
The thermal yield of particle $i$ in the fireball is given by:
\begin{equation}
\label{yieldT}
\langle N_i\rangle_T = Vg_i \int \frac{d^3p}{(2\pi)^3} \exp \left( -\frac{\sqrt {p^2+m_i^2}}{T}\right)  ,
\end{equation}
where all  chemical potentials were set to zero, as relevant for the beam energies at the  LHC considered here. We  also introduce  the density of particle $i$,  as  $n_i=\langle N_i\rangle_T/V$. 

To get the  total multiplicity of a hadron $i$ one has to add  resonances decaying to particle species $i$
\begin{equation}
\label{yieldD}
\langle N_i\rangle
= \langle N_i\rangle_T + \sum_j Br(j\rightarrow i) \langle N_j\rangle_T, 
\end{equation}
where $Br(j\rightarrow i)$ is the decay branching ratio of resonance $j$  to particle $i$. To include the width $\gamma_j$ of a resonance $j$ 
with mass $m_j$ in Eq.~\eqref{yieldD},  its thermal yield is calculated from, 
\begin{equation}
\label{yieldR} 
\langle N_j\rangle_{T}
=  V g_j\int \frac{d^3p}{(2\pi)^3}\int_{m_{th}}^\infty \frac{dM}{2 \pi} e ^{-\sqrt {p^2+M^2}/T} \, \Gamma_j(M), 
\end{equation}
where $\Gamma_j(M)$ is the relativistic Breit–Wigner distribution function
\begin{align}
		      \Gamma_j(M) = \frac{4 M^2 \gamma_j}{(M^2-m_j^2)^2 + M^2 \gamma_j^2},
\end{align}
and $m_{th}$ is the value of the invariant mass at threshold   for a give  decay channel of the resonance.  

The results,  summarized in 
Eqs.~(\ref{yieldT}-\ref{yieldR}),  extended to quantum statistics,  constitute the HRG model in the GC ensemble   with all chemical potentials $\vec\mu =0$, i.e. the  system is charge neutral. 
Such a  model has been  applied to quantify thermalization and particle production  in most central heavy-ion collisions at the LHC.
However, it is already well established,  that the GC model can overestimate the  yield of particles. 
This is particularly  the case if data  are taken in low multiplicity events where the fireball volume is small or at low collision energies where the temperature is low. 
In such cases, a thermal description requires  exact implementation of charge  conservation which is usually  described in  the C-ensemble~\cite{Hagedorn:1971mc, Hagedorn:1984uy,Cleymans:1990mn,Cleymans:1998yb,Hamieh:2000tk,Ko:2000vp,BraunMunzinger:2003zd}.

In the following, we focus on  exact strangeness conservation and assume that all other quantum numbers are conserved on average in the GC ensemble 
with $\vec \mu=0$. 
As mentioned earlier, this   is a good approximation for the thermal description of the collision fireball produced at LHC energies.  
In this case, the canonical ensemble with exact implementation of strangeness conservation and total strangeness $S=0$
is  achieved by introducing a  delta function under the trace in Eq.~\eqref{eq1},

\begin{equation}
Z^C_{S=0} = \textrm{Tr}\left[ e^{-H/T}\delta_{(S,0)}\right]  .
\end{equation}
For a non-interacting Hamiltonian and after performing the Fourier  decomposition  of the  delta function,   the canonical partition function can be written in  the following integral representation
 \begin{equation}
Z^C_{S=0}=\frac{1}{2\pi}
       \int_{-\pi}^{\pi}
    d\phi~ \exp{\left(\sum_{s=- 3}^3S_se^{is\phi}\right)},
    \label{eq6}
\end{equation}
where  $S_s= \sum_k z_{k,s}$  and the sum is taken  over all particles and
resonances that carry   strangeness $s$.   The one-particle  partition function is defined, as  
$z_{k,s}=V_C n_k^s(T)$  with the particle  density $n_k^s(T)$ as  in Eq. \eqref{yieldT}. In view of $\vec \mu=\vec{0}$ the thermal phase space of particles is the same as for antiparticles, i.e. $S_s=S_{-s}$. 
The volume  $V_C$ is the volume where  exact strangeness conservation $S=0$ is fulfilled. 

The strangeness canonical partition function in  Eq.~\eqref{eq6} can  also be expressed as  a series of  Bessel functions \cite{BraunMunzinger:2001as,BraunMunzinger:2003zd,Cleymans:1990mn,Hamieh:2000tk}, 
\begin{eqnarray}
Z^C_{S=0}&=& \sum_{n,p=-\infty}^{\infty}
 I_n(S_2) I_p(S_3) I_{-2n-3p}(S_1). 
\label{eq7}
\end{eqnarray}

 The yields of strange particles are usually measured in a given  acceptance region,  often corresponding to a restricted region in rapidity space. 
 Consequently,  the  single-particle  partition  functions $z_{k,s}$  should be  split   into two parts, $z_{k,s}=z_{k,s}^A+z_{k,s}^R$,  where
 $z_{k,s}^A$ stands for a particle  in the acceptance and  $z_{k,s}^R$      for those outside the acceptance region \cite{Braun-Munzinger:2020jbk}. 
 Furthermore, 
 we parameterize $z_{k,s}^A=V_A n_k^s(T)$, where $V_A$ is the volume in the acceptance window. 

For the calculation of the  mean multiplicity   of particle  $k$ carrying strangeness $s$ in the acceptance,  one introduces 
the auxiliary parameter  $\lambda_k^ A$ in Eq.~\eqref{eq6} by replacing $z_{k,s}^A \to \lambda_k^ A z_{k,s}^A$.   
The resulting mean multiplicities in the canonical ensemble in a given experimental acceptance  are obtained from
\begin{eqnarray}\label{NCML}
\langle N_k^s\rangle_A &=& \left.\lambda_k^ A{\frac{\partial\ln Z_{S=0}^C} {\partial\lambda_k^A}}\right|_{\lambda_k^A=1},
\end{eqnarray}
in the  following form: 

\begin{align}
\label{equ9}
\begin{split}
\langle N_k^s\rangle_A =& V_A \, n_k^s(T) \,  \frac{1}{Z_{S=0}^C} \times \\
& \sum_{n,p=-\infty}^{\infty} \, I_n(S_2) I_p(S_3) I_{-2n-3p- s}(S_1).
\end{split}
\end{align}
It is thus clear, that global and exact strangeness conservation constrained to $S=0$ in the full phase-space 
influences  yields of strange particles in the  experimental subspace. The yields are quantified 
by the temperature and  two volume parameters: the volume  of the system in the acceptance  $V_A$ and the correlation volume $V_C$ of  global strangeness  conservation which appear in the arguments of the Bessel functions. Furthermore, to get  the total multiplicity of a given strange particle  in C-ensemble, the decays of resonances have to be added  as in Eq.~\eqref{yieldD}, albeit with thermal contribution of strange  resonances  calculated from  Eq.~\eqref{NCML}.

The first two terms in  Eq.~\eqref{equ9} constitute the yields of strange particles in the GC ensemble as introduced in 
Eq.~\eqref{yieldT}, whereas the last terms describe the  strangeness  canonical corrections. To identify their  contribution  and dependence on particle strange quantum number $s$, one considers only the leading terms,  which correspond to $p=n=0$ in the series in Eqs.~(\ref{eq7}) and (\ref{equ9}), 
\begin{align}
\label{equ10}
\langle N_k^s\rangle_A \simeq V_A \, n_k^s(T) \, \frac{I_{s}(S_1)}{I_{0}(S_1)}.
\end{align}
Thus, the ratio of ${I_{s}(S_1)/ I_{0}(S_1)}$ is just the suppression factor which  decreases in magnitude   with increasing  $s$ of  hadrons  and with decreasing thermal phase-space occupied by strange particles,  as described by the argument  $S_1$  of the Bessel functions.  A decrease of $S_1=V_C\sum_k n(k,T)$ is  due to decreasing $T$,  thus also $\sqrt s$,  or decreasing $V_C$ which scales with charged particle density. These are the main properties of strangeness canonical suppression that have been introduced  \cite{Hamieh:2000tk} to describe  thermal production of multi(strange) hadrons   in heavy-ion collisions. 

In the following, we apply the HRG model described  above in the  C-ensemble to quantify   production of (multi-)strange hadrons and their 
behavior with charged particle multiplicity  as observed by the ALICE collaboration in different colliding systems and collision energies at the LHC.  To this end we also correct the HRG model with a more complete implementation of interactions within the S-matrix formalism as described in the next section.

\section{S-matrix and HRG}
\label{sec3}

Interactions among hadrons modify the density of states (DOS) of a thermal system and hence the thermal abundances of hadron states~\cite{Dashen:1969ep,Venugopalan:1992hy,Lo:2017sde,Lo:2020phg}.
In the scattering matrix (S-matrix) formulation of statistical mechanics, 
an effective spectral function $B(M)$ describing the DOS can be computed from the S-matrix~\cite{Dashen:1969ep}: 

\begin{align}
    \label{eq:B}
	B(M) =  \frac{1}{2} \, {{\rm Im} \, \rm Tr} \, \left[ \, S^{-1} \frac{\partial}{\partial M}  S - \left(\frac{\partial}{\partial M} S^{-1} \right) S \, \right].
\end{align}
where $M$ is the center-of-mass energy of the system.
The quantity $B(M)$ summarizes the various interactions among the scattering channels. 
For the simple case of a single-channel, two-body scattering, the phase shift $\delta(M)$
uniquely identifies the DOS due to the presence of an interaction.
For example, when the interaction is dominated by a single resonance of mass
$m_{\rm res}$ and width $\gamma(M)$, the resonant phase shift can be written as

\begin{align}
\label{eq:ps1}
    \delta_{\rm res}(M) = \tan^{-1} \frac{M \, \gamma(M)}{m_{\rm res}^2-M^2}.
\end{align}
The effective spectral function $B$ assumes the standard Breit-Wigner form 
upon neglecting the energy dependence of the numerator $ \left(M \, \gamma(M)\right) \rightarrow
\left(M \, \gamma_{\rm bw}\right) $ in Eq.~\eqref{eq:ps1}, such that

\begin{align}
	\label{eqn:smat2bw}
	\begin{split}
	B_{\rm res}(M) &= 2 \frac{d}{d M} \delta_{\rm res}(M) \\
		       &\approx 2 M \times \frac{2 M \gamma_{\rm bw}}{(M^2-{m_{\rm res}}^2)^2 + M^2 \gamma_{\rm bw}^2}.
	\end{split}
\end{align}
Note that the normalization condition
\begin{align}
	\int_{m_{th}}^{\infty} \frac{dM}{2 \pi} \, B_{\rm res}(M) = 1
\end{align}
is satisfied provided that 

\begin{align}
\delta_{\rm res}(\infty)- \delta_{\rm res}(m_{th}) = \pi
\end{align}
for a threshold energy $m_{th}$.

Furthermore, when the empirical phase shift from a partial wave analysis (PWA)
of the scattering experiment is used for the calculation of $B(M)$,
both resonant and non-resonant interactions are correctly incorporated, 
and the result becomes insensitive to the choice of parameters in an individual model.

As energy increases, new interaction channels open  and the scattering becomes inelastic. 
The prescription of Eq.~\eqref{eq:B} remains valid, 
but the S-matrix should now be formally understood as a matrix acting in the
open-channel space, i.e. an $N_{\rm chan.} \times N_{\rm chan.}$ matrix. 
The trace operation (${\rm Tr}$) in Eq.~\eqref{eq:B}, originated from the thermal trace in constructing the
partition function, enforces a summation over the $N_{\rm chan.}$ channels.
The effective spectral function describing a multichannel system reads~\cite{Fernandez-Ramirez:2018vzu,Lo:2017sde,Lo:2020phg}

\begin{align}
    B(M) =  \sum_{a=1}^{N_{\rm chan.}} \, B_a(M),
\end{align}
where

\begin{align}
    \label{eq:Bi}
    B_a(M) &= \frac{1}{2} \, {\rm Im}  \left[ \, S^{-1} \frac{d}{d M}  S -
    \left(\frac{d}{d M} S^{-1} \right) S \, \right]_{aa}.
\end{align}
Here $[ \ldots ]_{aa}$ means the $a-$th diagonal matrix element.

The channel-specific spectral function $B_a(M)$ describes the (energy-dependent)
component of the full $B(M)$ when projected into a specific interaction channel $a$. 
It generalizes the standard prescription of a 
branching ratio ($Br(X \rightarrow a)$) in Eq.~\eqref{yieldD} 
and reflects the energy dependence from resonant and non-resonant interactions.
The channel yield, e.g., from a resonance decaying into multiple final states,
can be readily computed via
\begin{align}
n_a(T) =  \int_{m_{th}}^\infty \frac{dM}{2\pi} B_a(M) \, n^{(0)}(T,M),
\end{align}
where $n^{(0)}$ is the ideal gas formula for the particle density.

In this study we focus on computing the hadron yields of protons and
$\Lambda+\Sigma^0$-baryons. 
For the protons, we follow previous studies and employ the empirical phase shifts from the GWU/SAID PWA on the $\pi N$ scattering~\cite{Workman:2012hx}. 
In addition, we implement a $\pi \pi N$ background contribution based on
the LQCD computation of the baryon charge-susceptibility $\chi_{BQ}$.
For the strange ($\vert S \vert =1$) baryon system, we 
employ an existing coupled-channel model involving $\bar{K} N$, $\pi \Lambda$, and $\pi \Sigma$ interactions. 
The hyperon model allows to calculate not only the thermal yields of strange baryons,
$\Lambda$'s and $\Sigma$'s, 
but also  to quantify the additional contribution of protons from the strange baryons.
We emphasize that the S-matrix scheme for $\vert S \vert = 1$ hyperons is based on Ref.~\cite{hyp_ccm} and 
that the same scheme was used to compute the corresponding contribution to $\chi_{BS}$ in~\cite{Fernandez-Ramirez:2018vzu}.
 No extra tuning was done in the $\vert S \vert = 1$ sector in this calculation.

\subsection{Composition of hadron yields}
Hadron yields predicted by a thermal model can be generally separated into two parts: 
a purely thermal yield determined by the freezeout parameters $T_f$ and
$\vec \mu_f$,
and a contribution which describes the multiparticle interaction involving the hadron species under study.
The latter can be further classified according to the conserved quantum numbers of the interaction channel.
For example the total proton yield can be written as

\begin{align}
    \langle p \rangle = \langle p \rangle_{\rm th} + \langle p \rangle_{N^*} +
    \langle p \rangle_{\Delta} + \langle p \rangle_{\bar{K} N} + \ldots.
\end{align}
where the notions $N^*$, $\Delta$, etc. stand for the general quantum numbers ($B=1; I=1/2, 3/2$) rather than referring to a specific resonance state.

The proton yield from a given interaction channel can be related to its thermal abundance. 
A further simplification enters when we focus on the freezeout conditions for describing hadron production in central nucleus-nucleus collisions
at LHC energies: the freezeout chemical potentials are
practically zero, and by isospin symmetry all isospin charge states contribute equally, giving

\begin{align}
    \begin{split}
        \langle p \rangle_{N^*} &= \frac{2}{3} \, \langle N^*_{Q=0} \rangle +
        \frac{1}{3} \, \langle N^*_{Q=1} \rangle \\
        &\approx \frac{1}{2} \, \langle N^* \rangle
    \end{split}
\end{align}
where $\langle N^* \rangle = \langle N^*_{Q=0} \rangle + \langle N^*_{Q=1} \rangle$ 
is the total thermal yield of $N^*$. The analysis proceeds similarly for protons emanating from $\Delta$'s and other channels composed of a single nucleon.

\begin{figure*}[ht]
\centering
\includegraphics[width=0.49\linewidth]{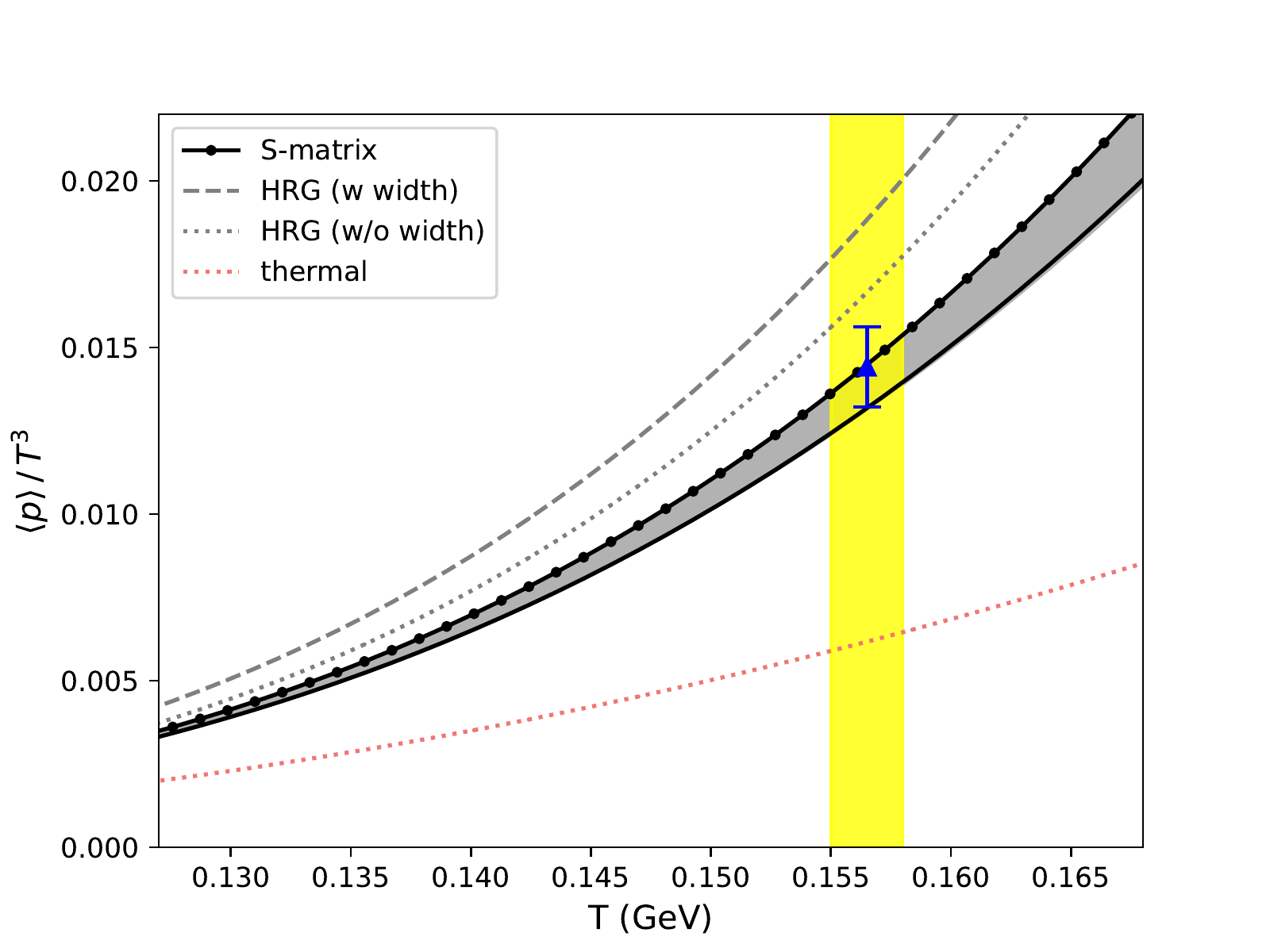}
\includegraphics[width=0.49\linewidth]{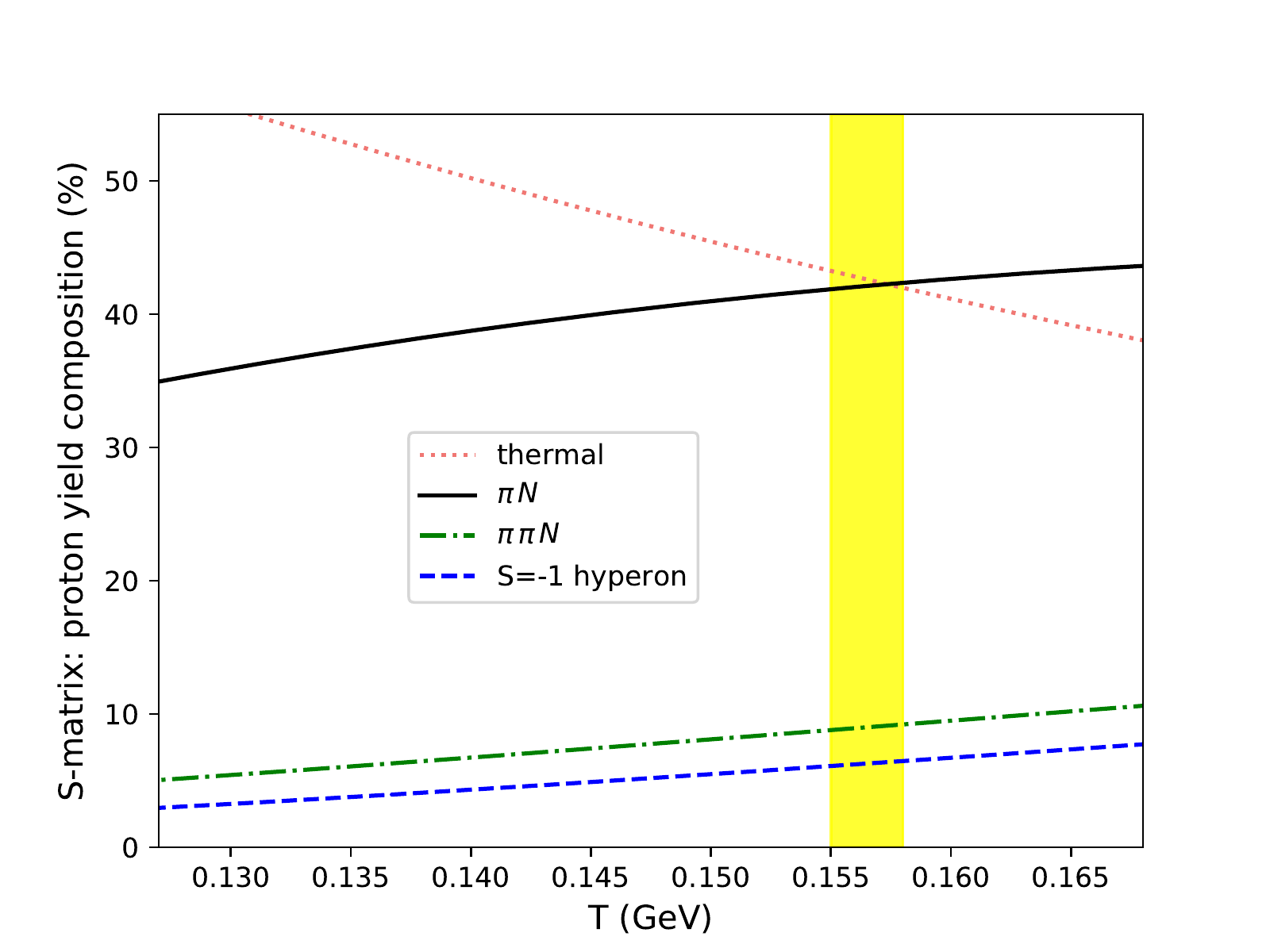}
        \caption{
        The proton yield density (left) and a composition analysis (right)
        computed within the S-matrix formulation.
        The corresponding results from the HRG scheme, implemented with and without resonance widths, are also shown.
        For the S-matrix prediction: 
        solid black points correspond to including the full list of interactions, 
        and the black line represents the partial result
        accounting only for the elastic and quasi-elastic scatterings, i.e.
        without the $\pi \pi N$ background contribution constrained by the LQCD result for $\chi_{BQ}$. 
        The blue triangle depicts the proton yield density calculated from the 
        proton yield data measured by the ALICE collaboration for the most
        central Pb-Pb collisions (at $2.76$ TeV) and the fireball volumes in Eq.~\eqref{eq:volume}.
        The error bar represents only the experimental errors of the yield data.
        }
\label{fig:fig2}
\end{figure*}

\begin{figure*}[ht]
\centering
\includegraphics[width=0.49\linewidth]{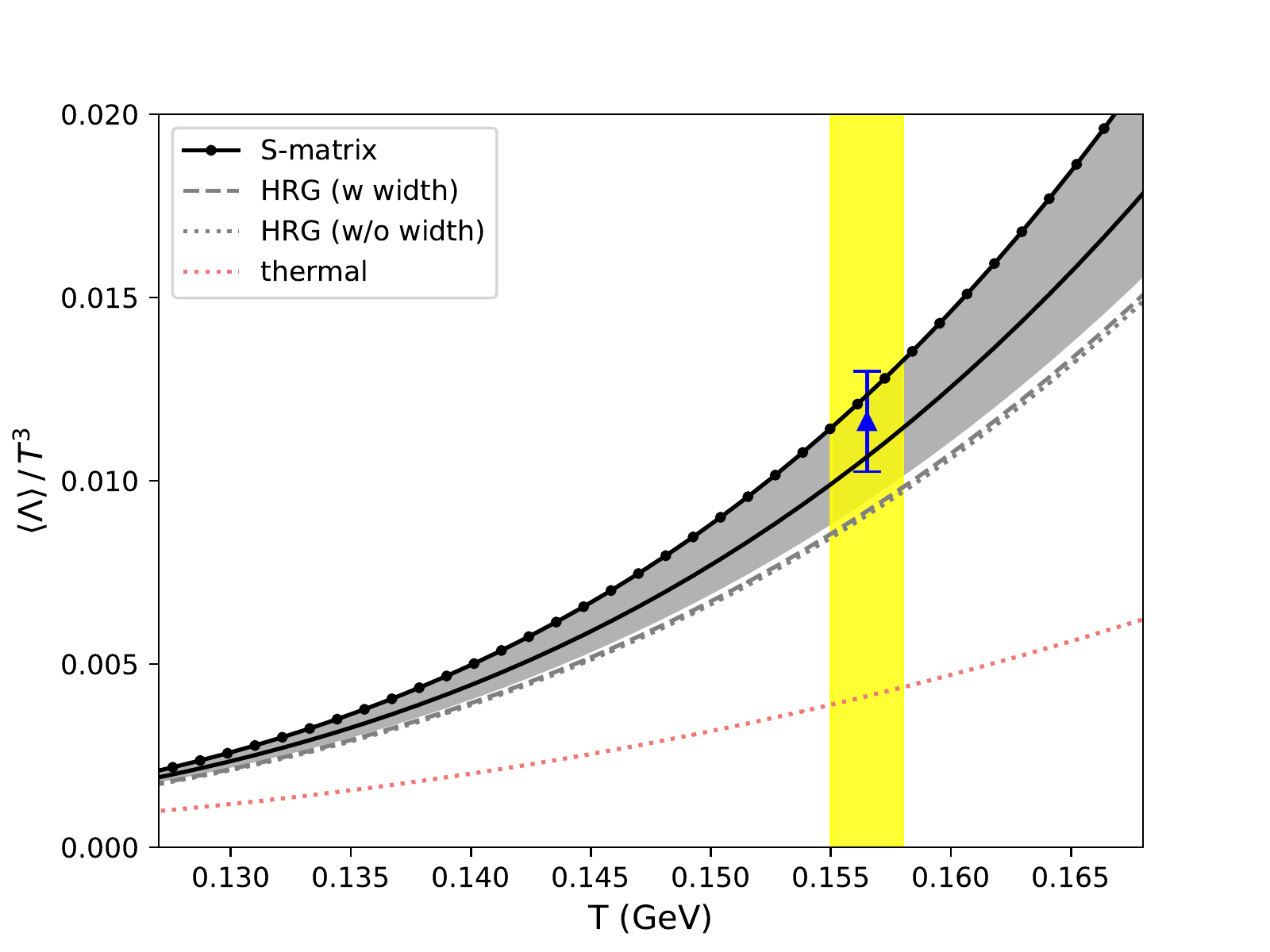}
\includegraphics[width=0.49\linewidth]{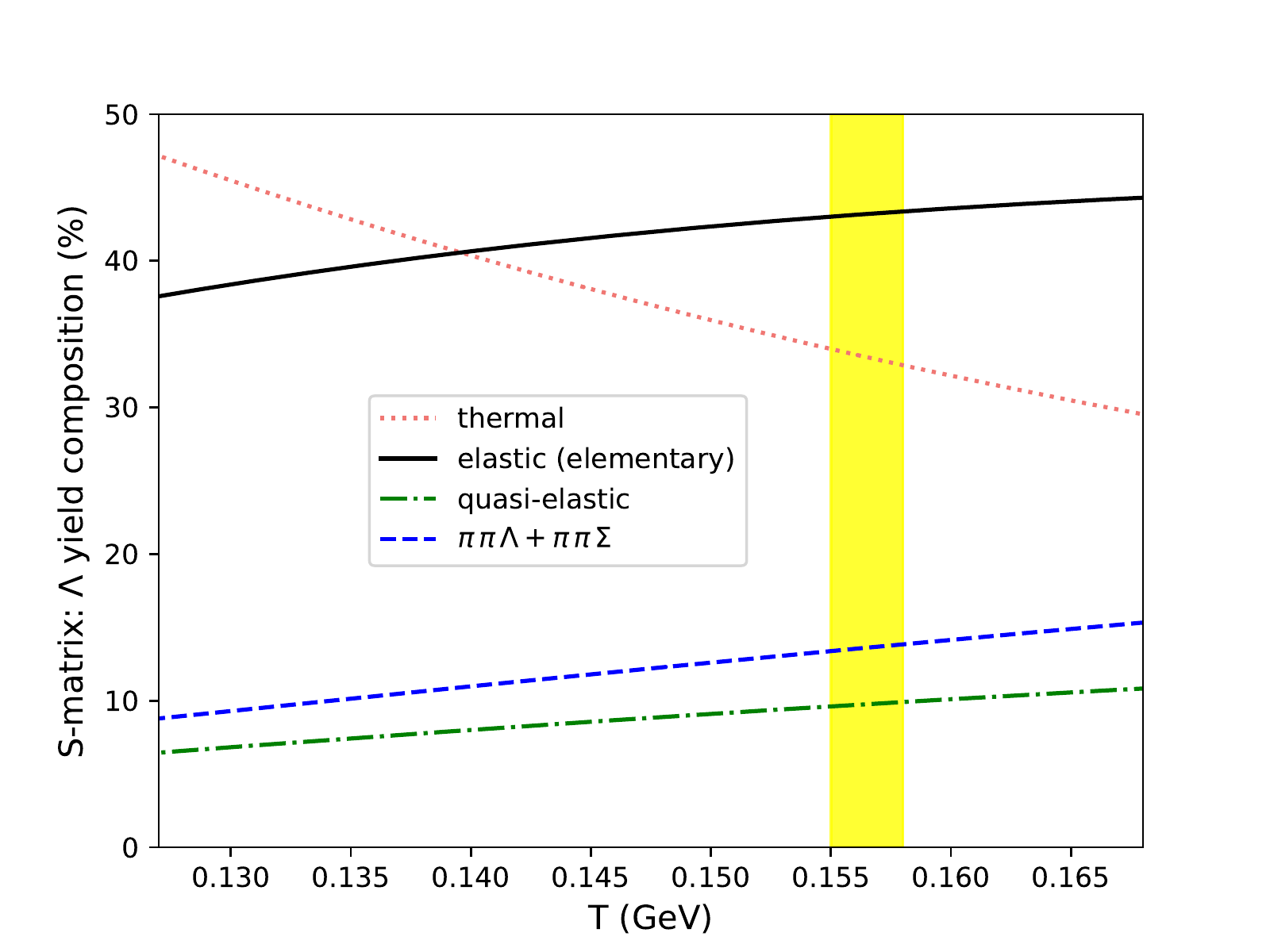}
        \caption{
        Similar to Fig.~\ref{fig:fig2} but for the $\Lambda+\Sigma^0$-baryon yield density. 
        The shaded region (in gray) shows the range of predictions by the S-matrix scheme at different levels of approximation: 
        from accounting only for the elastic scattering of elementary states (bottom), adding the quasi-elastic scatterings (black solid line), to including the full list of interactions (black points).
        See Table~\ref{tab1} for the list of interaction channels in the multichannel hyperon model.
        }
\label{fig:fig3}
\end{figure*}

\begin{figure}[ht]
\centering
\includegraphics[width=0.99\linewidth]{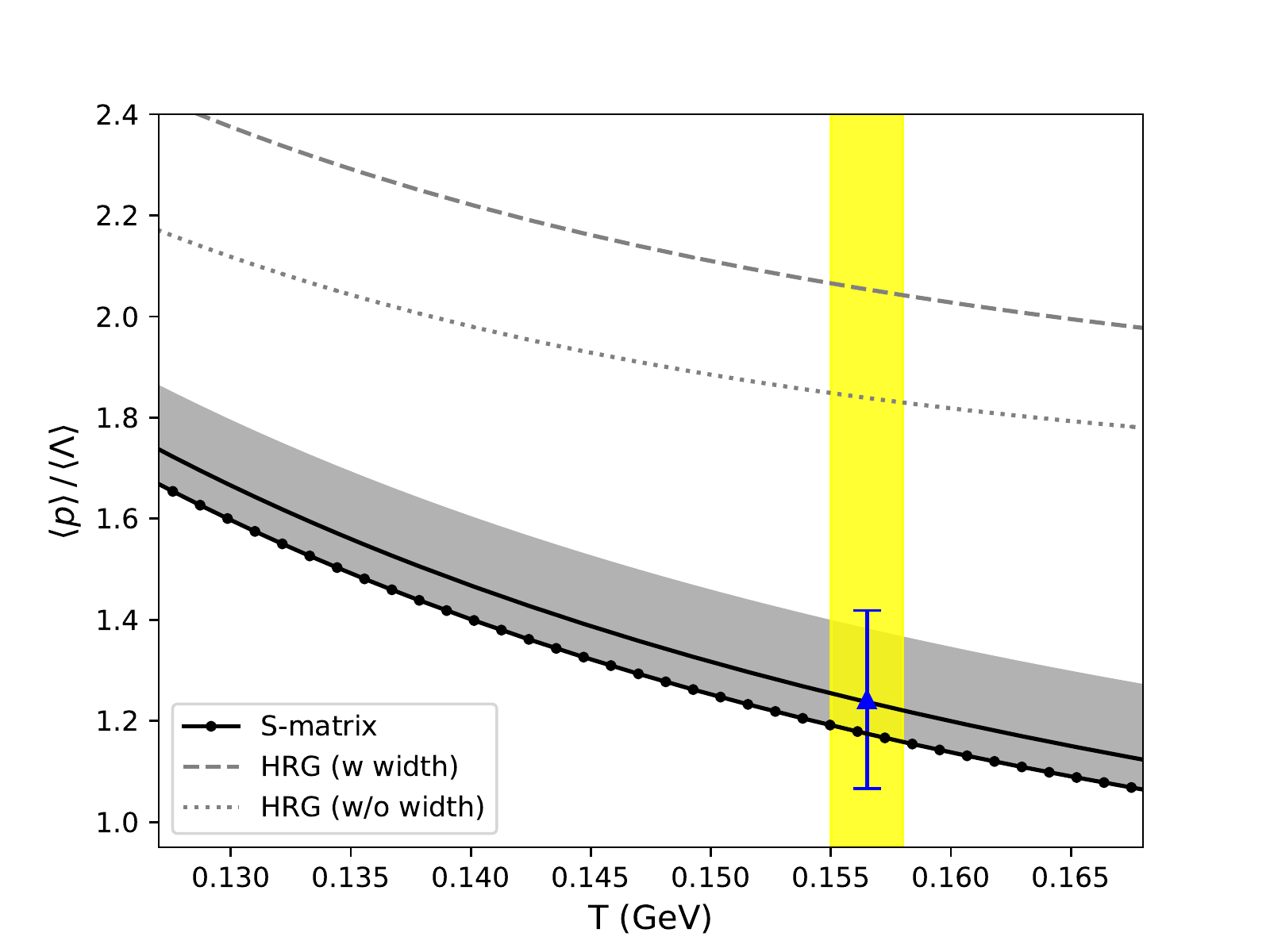}
	\caption{
	The ratio of the proton yield to that of $\Lambda+\Sigma^0$-baryons.
	The predictions of the S-matrix scheme at various levels of approximation (see text for conventions) and those from the HRG scheme (with and without width) are shown.
    The blue triangle depicts the experimental value measured by the ALICE collaboration for the most central Pb-Pb collisions (at $2.76$ TeV).
        }
\label{fig:fig4}
\end{figure}

We also compute the proton yields coming from the ($\vert S \vert=1$) strange baryons system.
The relevant channels are given by entries $(1, 6, 7, 8, 11-14)$ in Table~\ref{tab1}.
In particular, quasi-two-body states like $\pi \Lambda(1520)$ can
contribute via the channel weights $B_{13, 14}$, 
after further multiplication by an inherent branching ratio ($\approx 45 \%$ to protons) of the resonance $\Lambda(1520)$~\footnote{
Generally, channels involving $\Sigma(1385)$ and $\Lambda(1520)$ in the final states are multiplied by an inherent branching ratios, $(87, 12, 0 ) \, \%$ and $(10, 43, 45) \, \%$ respectively, when computing the yields of ($\Lambda$, $\Sigma$, protons).}.
However, the dominant contribution comes from the $\bar{K} N$ ($B_{1}$) channel, where no further manipulation is required.
The magnitude of $\langle p \rangle_{hyp.}$ from all hyperon channels is consistent
with the HRG estimate based on branching ratios, 
and amounts to $\approx 6 \%$ of the total proton yield at $T = 156.5$ MeV.

In Fig.~\ref{fig:fig2} we show the proton density (normalized to $T^3$) as a function of temperature. 
As noted in previous studies, implementing the essential features of the empirical $\pi N$ scattering, 
i.e. the widths of resonances and the non-resonant interactions, 
via the S-matrix formalism leads to a reduction of the proton yield relative to the HRG baseline. 
Including the protons from strong decays of $\vert S \vert =1$ hyperons does not alter this conclusion.

An estimate of the contribution by the $\pi \pi N$ background,
based on LQCD results on the baryon-charge susceptibility $\chi_{BQ}$, is also displayed. 
This gives $\approx 8.6 \%$ of the total proton yield at the LHC freezeout conditions.
Note that it is an independent source of protons compared to the ones from the
hyperons as the latter do not contribute to $\chi_{BQ}$ in an isospin-symmetric system~\cite{Lo:2017lym}.
A dissection of the composition of proton yield in the full S-matrix scheme is
shown in Fig.~\ref{fig:fig2} (right). 

We perform an analogous study for the $\Lambda+\Sigma^0$ yields based on the multichannel hyperon model. 
The result is shown in Fig.~\ref{fig:fig3} (left). 
Here it is convenient to classify the interaction channels into three groups:

\begin{table}[t]
    \begin{minipage}{0.48\textwidth}
    \centering
\begin{tabular}{|c|c|c|c|c|c|}
  \hline
        channel & elastic & channel   &  quasi-elastic  & channel & unitarity \\ \hline
        1       &  $\bar{K} N$        & 6   &  $\bar{K}^*_1 N           $   & 15  & $\pi \pi \Lambda $    \\ \hline
        2       &  $\pi \Sigma$       & 7   &  $[\bar{K}^*_3 N]_-       $   & 16  & $\pi \pi \Sigma  $    \\ \hline
        3       &  $\pi \Lambda$      & 8   &  $[\bar{K}^*_3 N]_+       $   &  &  \\ \hline
        4       &  $\eta \Lambda$     & 9   &  $[\pi \Sigma(1385)]_-      $   &  &  \\ \hline
        5       &  $\eta \Sigma$      & 10  & $[\pi \Sigma(1385)]_+       $   &  &  \\ \hline
                &                     & 11  & $[\bar{K} \Delta(1232)]_-     $   &  &  \\ \hline
                &                     & 12  & $[\bar{K} \Delta(1232)]_+     $   &  &  \\ \hline
                &                     & 13  & $[\pi \Lambda(1520)]_-  $   &  &  \\ \hline
                &                     & 14  & $[\pi \Lambda(1520)]_+  $   &  &  \\ \hline
\end{tabular}                                  
\caption{
    The list of interaction channels included in the coupled-channel PWA describing the $\vert S \vert = 1$ hyperon system 
        by the Joint Physics Analysis Center (JPAC) Collaboration~\cite{hyp_ccm}.
        Note that $\bar{K}^*$ is spin one and together with a nucleon can
        couple to spin $1/2$ (denoted $\bar{K}^*_1 N$) and spin $3/2$ (denoted
        $\bar{K}^*_3 N$). Subindices $\pm$ represent the higher and lower
        orbital angular momentum states which couple to a given partial wave.
        }
\label{tab1}
	\end{minipage}
\end{table}

\begin{itemize}

    \item[] Group I: elastic scatterings of elementary (ground state) hadrons (channel 1-5 in Table~\ref{tab1}).

    \item[] Group II: quasi-two-body states involving the resonances $K^*(892)$, $\Sigma(1385)$,
and $\Delta(1232)$ (channel 6-14 in Table~\ref{tab1}).

    \item[] Group III: dummy channels that collectively account for the missing inelasticity
    arising from channels not included explicitly (channel 15-16 in Table~\ref{tab1}).

\end{itemize}
Groups II and III should be formally understood as an effective treatment of three-body final states within the framework of an isobar decomposition, 
which is also compatible with the notion of effective elementary in thermal state counting~\cite{Dashen:1974jw,Dashen:1974yy}.

The $16 \times 16$ S-matrix is computed for the following partial waves: 
for $I=0$: $S_{01},~P_{01},~P_{03},~D_{03},~D_{05},~F_{05},~F_{07}$, and $G_{07}$ and for $I=1$: $S_{11},~P_{11},~P_{13},~D_{13},~D_{15},~F_{15},~F_{17}$, and $G_{17}$ cases. The subscripts specify the quantum numbers $(I, 2 \times J)$, where $I$ is isospin and $J$ is total spin.
These are included in computing the thermal yields of 
protons and hyperons.
        
Most channels in the Groups I and II have single-energy partial-wave data to fit, 
see~\cite{hyp_ccm} for details.
To distinguish between the different levels of approximation, we employ the following convention to present the results: 
The black points correspond to the full S-matrix scheme, including the full list of interactions.
The solid black lines are partial results accounting only for the elastic and quasi-elastic scatterings, i.e. Groups I and II.
The shaded region (in gray) shows the values spanned from including only Group I to the full list of interactions.

Unlike in the case of protons, the hyperon model predicts a higher $\Lambda+\Sigma^0$ yield than the HRG baseline.
The S-matrix approach entails the proper treatment of resonances and 
naturally incorporates some additional hyperon states beyond the listing of the 3- and 4-star PDG states~\cite{Tanabashi:2018oca}.
A table of the resonances identified in the PWA model is available from~\cite{Fernandez-Ramirez:2018vzu}.
The need for a stronger interaction strength has already been indicated 
by an analysis of the LQCD results of $\chi_{BS}$~\cite{Bazavov:2014xya}.

A composition analysis of the $\Lambda+\Sigma^0$ yield is shown in Fig.~\ref{fig:fig3} (right). 
The S-matrix scheme predicts that a substantial fraction of the hyperons comes from the quasi-two-body states and the 3-body backgrounds. 
The latter, constrained only by unitarity, represent the major uncertainty in the model.
Theoretical analysis of the amplitudes beyond the elastic, two-body limit is, as a rule, much more involved.
In particular, the proper treatment of these channels as genuine three-body states remains a challenging task, 
and is currently under active development~\cite{Hansen:2014eka,Briceno:2017tce,Mai:2017vot,Doring:2018xxx}.
To make progress we attempt to extract a phenomenological estimate of the
densities based on measured hadron yields and the fireball volume from a thermal model.

\subsection{Estimation of hadron densities}

As discussed, the LQCD results on $\chi_{BS}$ support an enhancement in the thermal yields of hyperons, 
while those on the $\chi_{BQ}$ suggests a suppression of thermal proton yield.
It is then of interest to investigate whether a particle yield analysis by a
thermal model would support these claims.

A key quantity to extract from a thermal model in the GC ensemble at a given temperature is the volume of the fireball. See Eq.~\eqref{eq:volume} and the discussion.
The volume is the same for all hadron species, 
and is obtained from a global fit to all hadrons.
Clearly, the volume thus obtained is not free from the influence of 
densities of protons and $\Lambda+\Sigma^0$-baryons, 
the latter we are trying to estimate. 
However, the value is most strongly determined by the most abundant hadrons in
the fireball, i.e., the pions.
Thus, constructing the hadron densities based on measured yields and the fireball volume 
gives a consistent estimate of the hadron densities by a thermal model.
With the volumes constructed for the most central Pb-Pb collisions (at $2.76$ TeV, averaged over the three data points with the highest multiplicities), 
the estimate for hadron densities are computed and are shown as blue triangles in Figs. \ref{fig:fig2} and \ref{fig:fig3}.

For protons the estimate agrees with the full S-matrix result.
Note the different origins of the error on the estimate and the gray band:
the former is due to the experimental error of the measured proton yield, 
the latter represents the effect of including or excluding the contribution from the $\pi \pi N$ backgrounds.
For the hyperons, the density estimate supports the trend predicted by the S-matrix approach: an increase in the hyperon yields compared to the HRG baseline.
The estimate substantiates the need for including quasi-two-body states 
(up to the black line in Fig.~\ref{fig:fig3} (left)) and 
in addition part of the three-body unitarity backgrounds.
The remaining discrepancy may be resolved 
by including other channels such as $K \, \Xi$ (and other multi-strange states), 
and an improved treatment of the three-body states.

It is also possible to construct observables that are independent of the fireball volume.
An interesting quantity to consider 
is the ratio of the yields of protons to that of $\Lambda+\Sigma^0$-baryons.
As discussed, the HRG scheme tends to overestimate the numerator (more so for the scheme with resonance widths) while underestimating  the denominator.
This results in a much larger value than the prediction of the S-matrix scheme as shown in
Fig.~\ref{fig:fig4}.
Evidently, the result obtained from the measured hadron yields favors the latter, 
demonstrating the robustness of the S-matrix scheme.
Further thermal model analysis of the measured particle yields will be presented in the next section.

\section{Model comparison with ALICE  data}
\label{sec4}

In the GC ensemble with $\vec \mu=\vec{0}$ and for constant temperature the yields of all particles 
per volume should be  clearly independent of $dN_{ch}/d\eta$. On the other hand, from  the previous section, it is clear that the density of charged particles in the C-ensemble exhibits a non-linear dependence on volume parameters.  Consequently, in a thermal model any dependence of charged  particle densities on $V$ 
can be a strong indication of  possible corrections due to  exact conservation of quantum numbers. Indeed, in the context of strangeness production at the LHC,  already   from a previous analysis~\cite{Sharma:2018jqf} it was clear that the canonical ensemble with exact strangeness conservation (CSE)
 is the best ensemble  for
describing the low multiplicity classes in p-p collisions at the LHC. 
Furthermore, the CSE was shown to connect better to the high multiplicity classes than 
the canonical ensemble with exact baryon, strangeness and
charge conservation~\cite{Sharma:2018jqf}. Thus,  in the following, we focus on only exact strangeness conservation and analyze data obtained by the  ALICE 
collaboration on (multi-)strange particle multiplicity in p-p collisions  at 7 TeV~\cite{ALICE:2017jyt} as well as
13 TeV \cite{Acharya:2019kyh},  
in  p-Pb collisions at 5.02 TeV~\cite{Abelev:2013haa,Adam:2015vsf} and in  Pb-Pb collisions
at 2.76 TeV~\cite{Abelev:2013vea,Abelev:2013xaa,Abelev:2013zaa}.
The thermal model calculations
are done using the latest version of THERMUS~\cite{Wheaton:2004qb}~\footnote{B. Hippolyte and Y. Schutz,
https://github.com/thermus-project/THERMUS.} {which is  further extended here   to account for   a more complete description of hadron interactions within the S-matrix approach. We consider the S-matrix corrections to proton and hyperon yields as introduced  in the previous section.} 

As a first step, we made a fit for each multiplicity bin 
by keeping the number of parameters to a minimum.
For the temperature  we are guided by the results from LQCD \cite{Bazavov:2018mes} and the recent HRG model analysis of ALICE data for central Pb–Pb collisions \cite{Andronic:2017pug,Andronic:2018qqt}  and allow only two choices:   $T_f$ = 156.5 MeV and $T_f$ = 160 MeV, 
to identify  
$T$-variation of particle yields.  
For the strangeness suppression factor $\gamma_s$ we 
work with the value $\gamma_s = 1$ as motivated by fits  in central Pb-Pb collisions, this is in contrast to~\cite{Das:2016muc,Vovchenko:2019kes}
where substantial deviations have been proposed.
The chemical potentials due to  conservation of   baryon number and   electric charge are    being set to zero as in the energy region of interest at the LHC,  particle-antiparticle symmetry is observed
with a good degree of accuracy. Thus, in the SCE  only two parameters remain, the volume of the system in the experimental acceptance $V_A$  and the canonical volume $V_C$ which quantifies  the range  of exact strangeness conservation.
An independent determination of the $V_C$ parameter remains an open issue.

\begin{figure*}[ht]
\centering
\includegraphics[width=\linewidth, height=13cm]{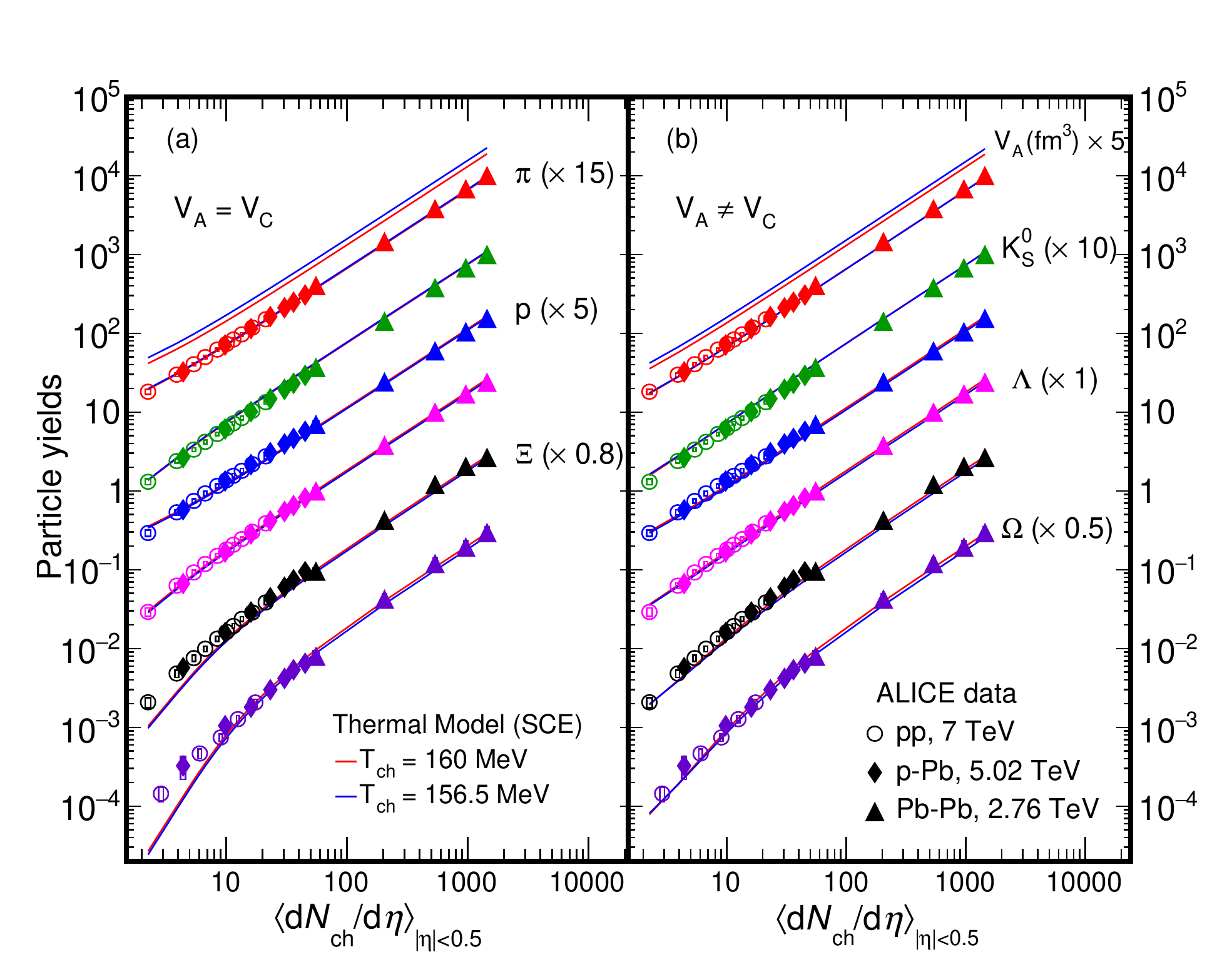}
	\caption{Left-hand figure: Yields for $V_A=V_C$. Right-hand figure: Yields for $V_A\neq V_C$,
The two	top lines are the fitted volumes $V_A$ (x5) in fm$^3$. The particle yields are indicated in the right panel together with the multiplicative factor used to separate the yields. The solid blue lines have been calculated for $T$ = 156.5 MeV while the solid red lines have been calculated for $T$ = 160 MeV. 
The values of the volumes used have been parametrized empirically   in Eqs.~(\ref{eq:volume}) and (\ref{eq:volumec}).
}
\label{fig:fig5}
\end{figure*}

For large multiplicities  we have  found  that $V_A \simeq  V_C$, 
therefore  we first put them equal
for all multiplicities. In Fig. \ref{fig:fig5} (left) we show the yields of hadrons calculated in the SCE for different charged particle multiplicities
$dN_{ch}/d\eta $ at fixed temperature and for a single volume  $V=V_A\simeq
V_C$. It is to be noted that,  the charged particle multiplicity  is measured
by ALICE collaboration in the pseudorapidity range $-0.5 < \eta <0.5$,  while the particle yields are measured in a rapidity interval 
$-0.9 < y < 0.9$.

 The SCE model is seen in Fig. \ref{fig:fig5} (left) to capture  basic properties of hadron yields data already with a single volume  parameter. For large $dN_{ch}/d\eta >  100$ all hadron yields, as well as  extracted $V$,   depend   linearly  on  charged particle rapidity density. However, 
for lower $dN_{ch}/d\eta$ this dependence is clearly non-linear for strange particles  due to strangeness canonical  suppression which increases with the strangeness content of particles. On the quantitative level, however, one can see  in  Fig. \ref{fig:fig5} (left), that using a single volume,  leads to too much suppression at small  charged particle multiplicities,  particularly for $S=-2$ and $S=-3$
baryons.
This result is consistent with the  previous observation, that a single volume canonical model implies   too strong  strangeness suppression in low multiplicity events~\cite{Adam:2015vsf,Vislavicius:2016rwi,Anielski:2014}. 

In general, strangeness conservation relates to the full phase-space whereas particle yields are measured in some acceptance window. Thus, the strangeness canonical volume  parameter  $V_C$ can be larger than the fireball volume $V_A$,   restricted to a given acceptance. To quantify ALICE data we have performed the SCE model fit  to data  with two independent volume parameters as shown in Fig. \ref{fig:fig5} (right). The resulting yields  exhibit  much better agreement with data by decreasing strangeness suppression at lower multiplicities due to  larger value of $V_C$ than $V_A$.  

\begin{figure}[htp]
\begin{center}
\includegraphics[width=0.5\textwidth,height=9cm]{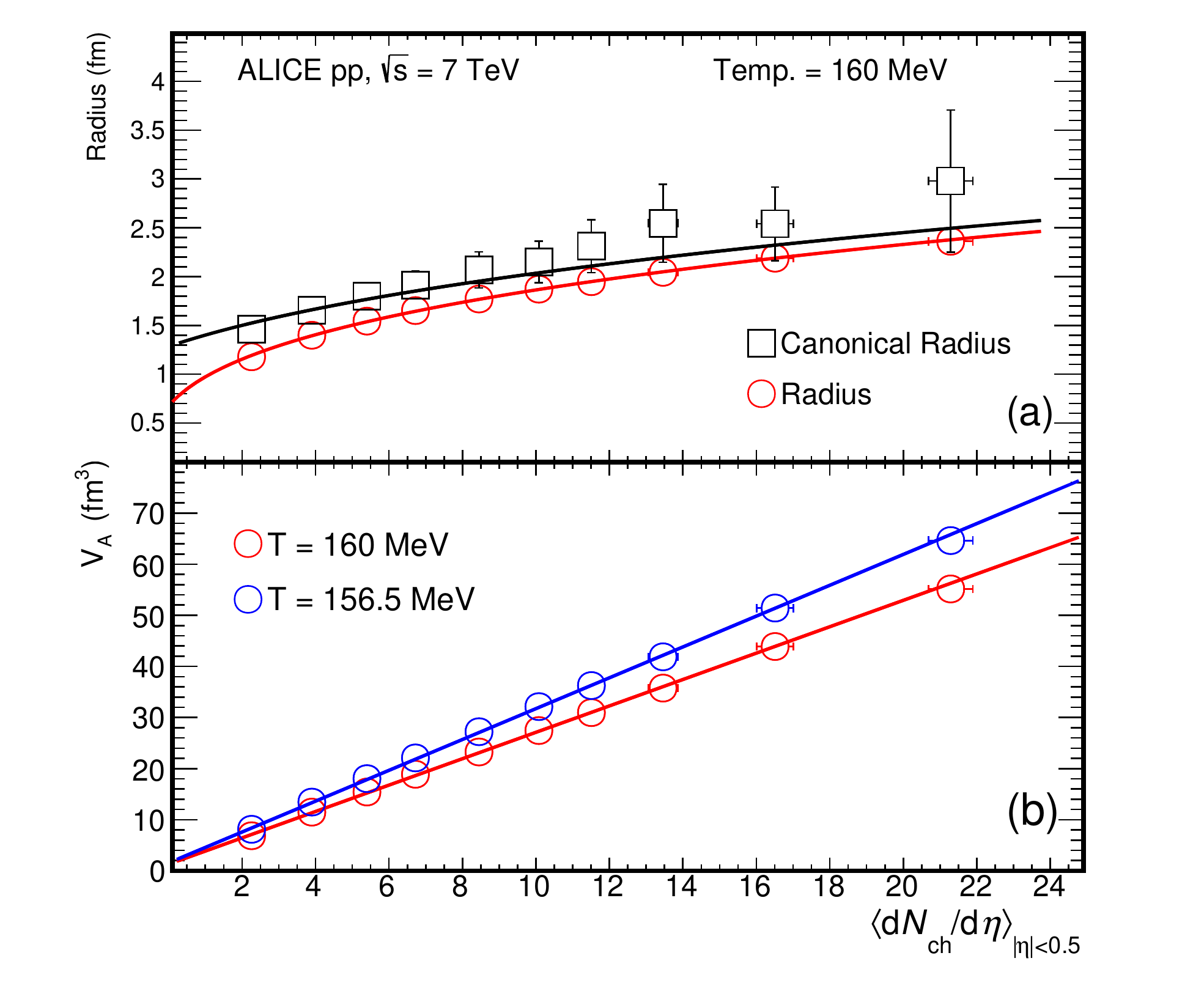}
\caption{
The upper panel
shows the radius (red points) as extracted from fits to p-p collisions at 7 TeV as a function
of the charged hadron multiplicity in mid-rapidity region using the SCE model 
with $T$ = 160 MeV and $\gamma_s$ =1.
The black points in the upper figure show the values of the canonical radius.
The lower panel shows the volume $V_A=4/3\pi R^3$ as a function of the charged particle multiplicity for two values of the chemical 
freezeout temperature, red points were obtained from 
$T$ = 160 MeV, blue points are for $T$ = 156.5 MeV.
}
\label{fig:fig6}
\end{center}
\end{figure}

The fitted volume parameters  are
shown in Fig.~\ref{fig:fig6} for $T$ = 160 MeV. Two features are  to be noted 
in this figure. First of all the overall volume $V_A$ can be determined fairly accurately and  increases linearly
with the charged-particle multiplicity. This strongly supports that the yields are directly proportional to the volume of the fireball   and  
agrees with one of the basic ingredients of the thermal model.
Second, the canonical volume $V_C$ differs from $V_A$, however the difference 
is not so well determined at larger $dN_{ch}/d\eta$ as it appears in a ratio
of Bessel functions which is already near   to its  asymptotic value. 
For small multiplicities,  however, the value of $V_C$  is clearly larger than $V_A$ leading 
to a reduced suppression of  strange-particle yields. 
The fits to $V_A$ and $V_C$  were  made for each multiplicity bin and can be well parametrized  
as linear functions of charged particle multiplicity:

\begin{align}
\begin{split}
    V_A &= 1.27 + 2.58 \times \frac{dN_{ch}}{d\eta} ~~\textrm{for}~T = 160~\textrm{MeV}\\ 
    V_A &= 1.55 + 3.02 \times \frac{dN_{ch}}{d\eta}
    ~~\textrm{for}~T = 156.5~\textrm{MeV}
\label{eq:volume}
\end{split}
\end{align}
The volume is slightly larger for $T$ = 156.5 MeV than for $T$ = 160 MeV so as to compensate for the smaller particle densities.
A possible form for the canonical volume is given by
\begin{align}
    \begin{split}
    V_C &= 8.87 + 2.64 \times \frac{dN_{ch}}{d\eta} ~~\textrm{~for~}~T = 160~ \textrm{~MeV}\\ 
    V_C &= 12.32 + 3.02 \times \frac{dN_{ch}}{d\eta}
    ~\textrm{for}~T = 156.5~\textrm{~MeV}
\label{eq:volumec}
\end{split}
\end{align}
These  parametrizations are shown in Fig.~\ref{fig:fig6} as lines and have been used in all our model comparisons   to  data.
All numbers in Eqs.~(\ref{eq:volume}) and ~(\ref{eq:volumec}) are in units of [fm$^3$]. 
We emphasize that the parametrizations of the volumes given in Eqs.~(\ref{eq:volume}) and (\ref{eq:volumec}) are purely empirical.
It is indeed interesting to see a linear dependence (in $dN/d\eta$) of the acceptance volume, which to a good accuracy is also independent of the collision system.

\begin{figure}
\includegraphics[width=0.5\textwidth,height=8cm]{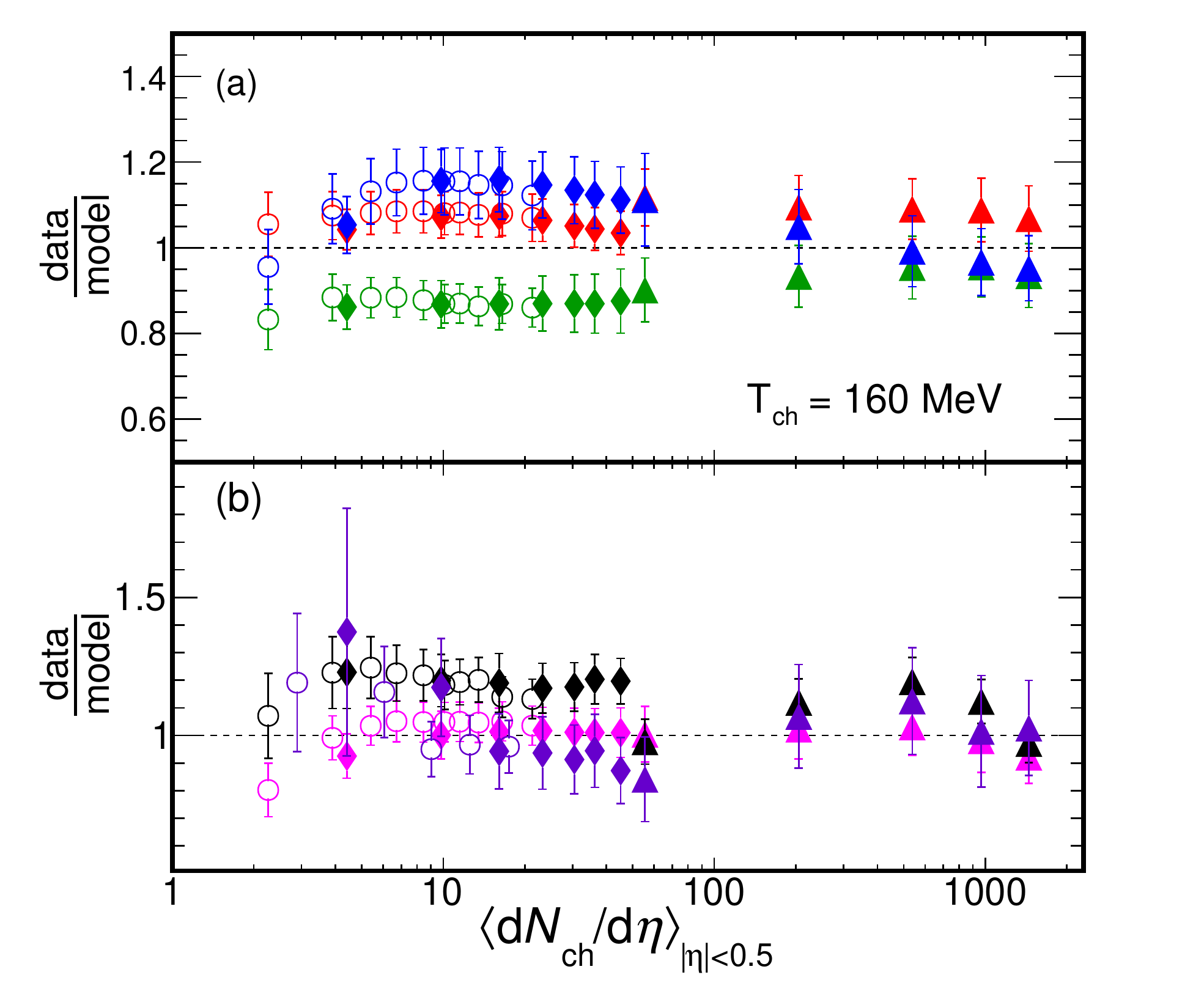}
\caption{
Data relative to model calculations. The upper panel shows the protons in blue,  the pions in red, the kaons in green. The lower panel shows the $\Xi$ in black, the $\Lambda$ in magenta and the $\Omega$ in purple. 
}
\label{fig:fig7}
\end{figure}

To appreciate the quality of  the SCE model description of ALICE data 
illustrated  in Fig.~\ref{fig:fig5} (right), we show in Fig.~\ref{fig:fig7} 
the ratio of data and the  model results. 
It can be seen that the model  prediction agrees quite well  with  the data up to  two standard deviations  for  all $dN_{ch}/d\eta$. 
The data on pion yields  are always slightly above the calculated points while the kaons are always below. This has implications for the kaon to pion
ratio discussed further below. This illustrates the pitfalls of showing ratios in the thermal model.
It is better to compare directly yields.

\begin{figure}[t]
\includegraphics[width=0.5\textwidth,height=8cm]{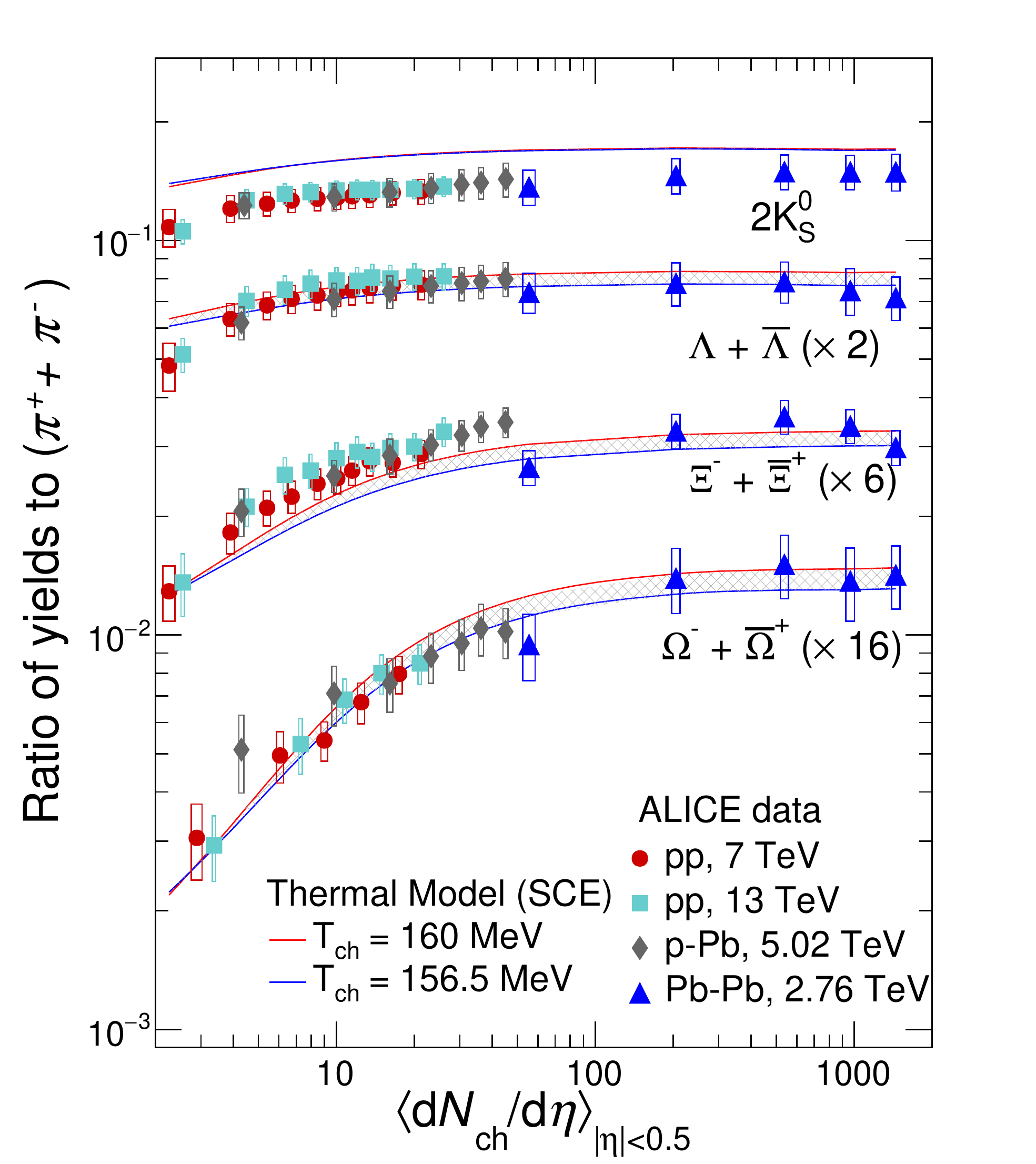}
\caption{ Ratios of yields of strange particles to pions versus charged particle multiplicity. The SCE model results were obtained using $V_A\neq V_C$ as explained in the text.}
\label{fig:fig8}
\end{figure}

The strangeness suppression effect and its CSE model description are particularly transparent when removing an overall  linear  dependence of  particle yields  on the fireball   volume $V_A$. This is e.g.  achieved by  plotting the ratio of strange particle and  pion yields,   as shown in Fig.~\ref{fig:fig8}. 
This ratio has been discussed prominently by the ALICE collaboration~\cite{ALICE:2017jyt} where 
a comparison with other model calculations was presented. The SCE  model introduced  here compares very favorably to the ones discussed in~\cite{ALICE:2017jyt}.
The underestimation of the 
pion yield is responsible for the larger discrepancy in the kaon to pion ratio as seen  in  Fig. \ref{fig:fig7}.

\begin{figure*}[ht]
\centering
\includegraphics[width=0.49\linewidth]{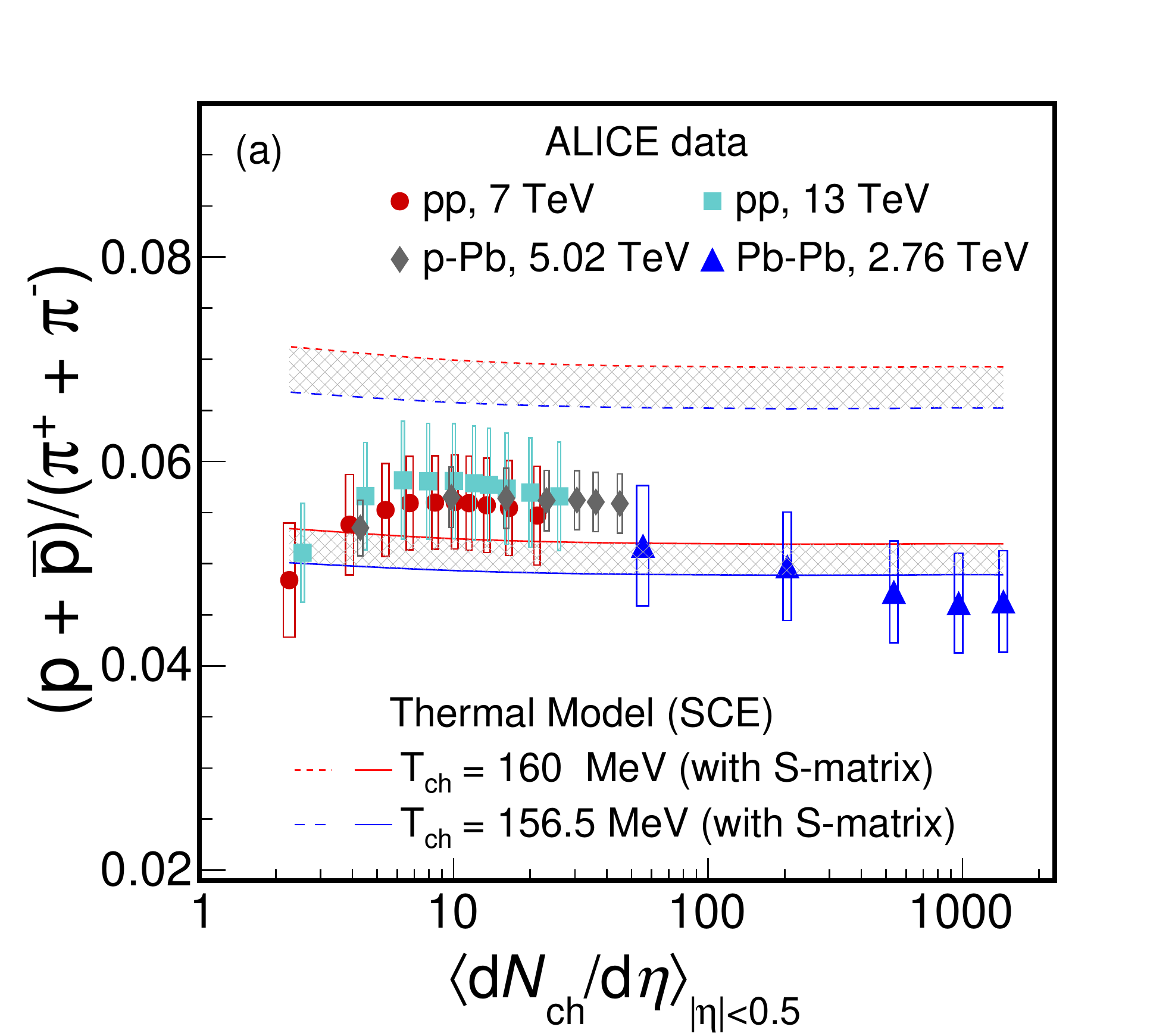}
\includegraphics[width=0.49\linewidth,height=7.7cm]{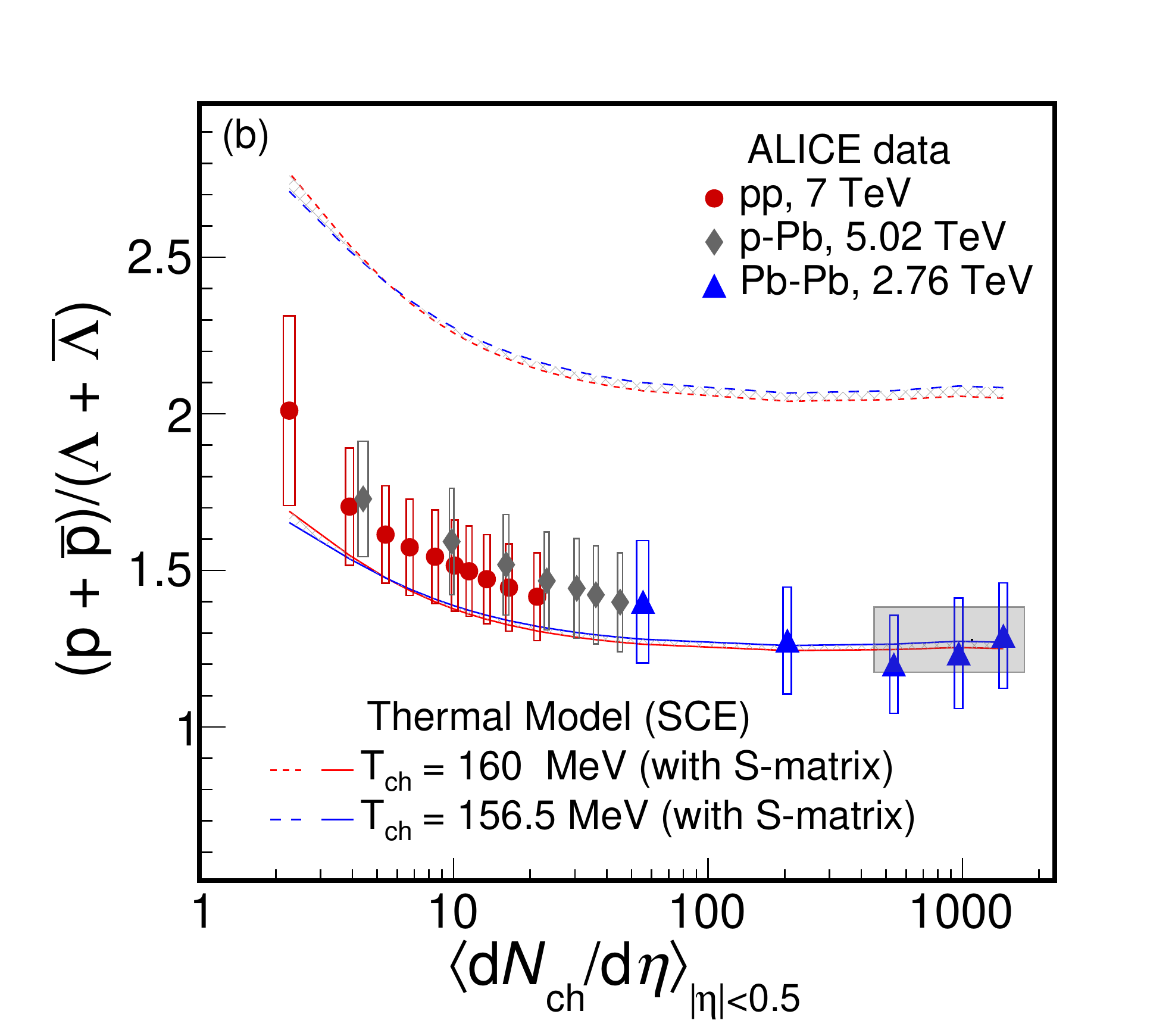}
\caption{
Left-hand figure: Proton to pion ratio showing the  importance of S-matrix corrections.
Right-hand figure: Proton to Lambda ratio showing the  importance of S-matrix corrections.
The lower lines take into account the S-matrix corrections, the upper lines are obtained without those corrections.
The shaded region (in gray) shows the theoretical predictions spanned by different stages of the S-matrix improvement: from including only elastic scatterings of ground state hadrons (upper limit) to the full list of interactions (lower limit). See text for details.
}
\label{fig:fig9}
\end{figure*}

\begin{figure*}[ht]
\begin{center}
\includegraphics[width=0.5\textwidth,height=8cm]{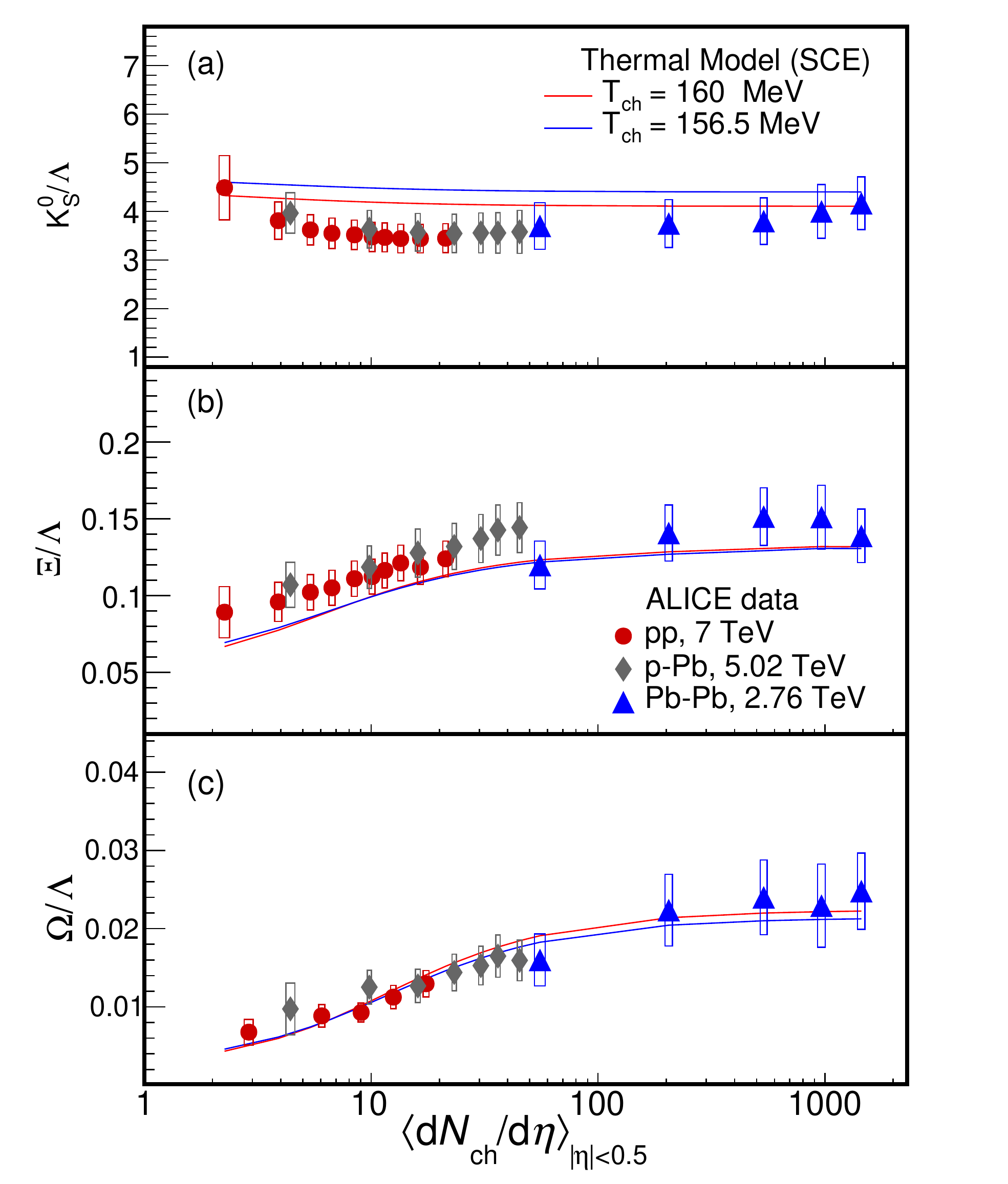}
\caption{Ratios of $K^0_S$
(upper) panel), $\Xi$ (middle panel) and $\Omega$ (lower panel) to $\Lambda$ yields. 
All lines were calculated taking into account S-matrix corrections as discussed in the text.
}
\label{fig:fig10}
\end{center}
\end{figure*}

The importance of the S-matrix description of proton and hyperon yields can be directly verified with data from the central Pb-Pb collisions 
by comparing measured $p/\Lambda$ ratio with the S-matrix results shown in Fig.~\ref{fig:fig4}. 
It is seen in this figure,  that data are indeed  well consistent with S-matrix predictions at $T_f\simeq T_c$,  indicating that particles are produced  at the phase boundary. Furthermore, such comparison can also be done for densities  of protons and hyperons as discuss 
in the context of Figs. \ref{fig:fig2} and \ref{fig:fig3} by using yields data and 
the fitted fireball volume for the central Pb-Pb collisions. 

To justify the S-matrix results for different charged particle densities we show  in Fig.~\ref{fig:fig9} the 
 $p/\pi$ (left) and  $p/\Lambda$ (right) ratios 
 as  functions of $dN_{ch}/d\eta$.  Both ratios are very sensitive to the S-matrix corrections which amount to a reduction of  proton yield by a factor of 0.75 and a 1.24 enhancement of the $\Lambda$ yield. The excellent agreement of data and the S-matrix values that have been already discussed  for  most central collisions are also verified for  lower charged particle multiplicities. The $p/\Lambda$  ratio is increasing with decreasing $dN_{ch}/d\eta$ with a strength  which is well consistent with the  SCE model. The thermal model without S-matrix corrections exhibits  large deviations from the  $p/\Lambda$  ratio data. 
 The $p/\pi$  is nearly multiplicity  independent since the SCE corrections to this ratio are small. Nevertheless, the  
$p/\pi$ ratio with  the  S-matrix corrections is well consistent within errors with the data at all $dN_{ch}/d\eta$. 

In Fig.~\ref{fig:fig10} we show  ratios of strange and multi-strange particles compared to the $\Lambda$ yield. The model results for the $\Omega/\Lambda$ ratio describe data very accurately while the two other ratios,  $K_S^0/\Lambda$ and $\Xi/\Lambda$,  are again within
one standard deviation. It is interesting to note, that in the SCE model the $K_S^0/\Lambda$ ratio  is independent of   $dN_{ch}/d\eta$. This is because the canonical suppression factor is the same for all $S=\pm 1$ mesons and baryons.   The results  shown in  Figs. \ref{fig:fig9} and \ref{fig:fig10} provide  further evidence  that  the S-matrix description of interactions in the proton and hyperon channels,  as well as,  that the strangeness suppression due to exact and global strangeness  conservation,  are  justified by data from the ALICE collaboration. 

\section{Summary and Conclusions}

We  have studied   the  influence  of  global  strangeness quantum number  conservation  on  strangeness production  in heavy-ion  and elementary collisions in a given acceptance region,  accounting  for  the  strangeness neutrality condition  formulated   in the hadron resonance gas (HRG) model in  the canonical ensemble.   We have focused on (multi-)strange baryon production yields in p-p, p-A and A-A collisions at the LHC energies and their behavior with charged-particle  multiplicity  at mid-rapidity,  as measured by the ALICE collaboration at CERN.  

To this end, the HRG model is augmented with the S-matrix corrections to the yields of protons and hyperons. 
The S-matrix calculation is based on the empirical phase shifts of $\pi N$ scattering, an estimate of the $\pi \pi N$ background constrained by Lattice QCD results of baryon-charge susceptibility, and an existing coupled-channel model describing the $\vert S \vert =1$ strange baryons.
It is demonstrated that an accurate description of the widths of resonances and the non-resonant interactions in a thermal model leads to a reduction of the proton yield relative to the HRG baseline (by $\approx 25\%$). Including the protons from strong decays of $\vert S \vert =1$ hyperons, which constitute $\approx 6\%$ of the total yields, does not alter this conclusion. 
Such a reduction is also crucial for resolving the proton anomaly in the LHC data.
For the hyperons, 
the S-matrix scheme predicts an increase in the $\Lambda+\Sigma^0$ yields relative to the HRG baseline by $\approx 23\%$.
This is consistent with the data from the ALICE collaboration.
Furthermore, the S-matrix prediction on the ratio of yields of proton to $\Lambda+\Sigma^0$ 
is in good agreement with the measured values by the  ALICE collaboration in Pb-Pb collisions in the events with the largest multiplicities ($dN_{ch}/d\eta$). The evolution of (multi-)strange baryons to $\Lambda$ yields with $dN_{ch}/d\eta$ calculated in the present thermal model in the C-ensemble 
follows the measured values within two standard deviations.

Comparing these experimental findings alongside the thermal model predictions 
(including the S-matrix corrections), 
it is evident that an accurate treatment of two-body scatterings 
and in addition the three-body interactions, 
at least those captured within the quasi two-body framework, 
is necessary for a satisfactory description of data.
A more realistic theoretical treatment of the three-body interactions 
may resolve the remaining background 
contribution.

A good description 
was obtained for the variation of the strangeness content in
the final state as a function of the
number of charged hadrons at mid-rapidity with  the same freezeout temperature $T_f\sim 156.5$ MeV   as calculated  previously from  data in the most central Pb-Pb collisions. This lends  further support that at LHC energies and independently of colliding system, the freezeout temperature   coincides  with the chiral crossover  as calculated in  LQCD. 

We have argued that the observed behavior of hadron yields at the  LHC  with  charged particle multiplicity can be explained  naturally in the thermal model. The temperature is linked to the collision energy and  is  independent of the colliding system. The fireball volume parameter   at mid-rapidity was found to scale linearly with $dN_{ch}/d\eta$. The observed increasing suppression of strange hadron yields  with decreasing $dN_{ch}/d\eta$  and its dependence on  their strangeness content,  was found to be qualitatively  consistent with predictions of the HRG model formulated  in the strangeness  canonical ensemble. An exact conservation of strangeness is to be imposed in the full phase-space rather than in the experimental acceptance at mid-rapidity. Consequently,  the correlation volume parameter where strangeness is exactly conserved was found to be larger than the fireball volume  at mid-rapidity.

The S-matrix formulation of statistical mechanics 
 accurately describes the measured hadron yields and supports
the interpretation of the LQCD results. 
This encourages its further adoption in analyzing the 
multi-strange baryon sectors and the light mesons.
The thermal yield data can guide future S-matrix modeling, which allows a more reliable assessment of the relevance of, for example, the extra Quark Model states~\cite{Loring:2001ky} in $\chi_{BS}$ and in higher order fluctuations.
Also, many Fock states considered in the current scheme contain multi-pion components. It is thus pivotal to investigate their influence on the thermal production of pions and kaons.
This requires an extensive analysis involving also the purely mesonic sector~\cite{Huovinen:2016xxq}, and will be explored in a future study \cite{new}.

\section{Acknowledgments}
P.M.L and K.R  acknowledge the support by the Polish National Science Center (NCN) under  Opus grant no.  2018/31/B/ST2/01663. 
K.R. also acknowledges partial support of the Polish Ministry of Science and Higher Education,  and fruitful discussions with 
Anton  Andronic, Peter  Braun-Munzinger, Bengt  Friman, Anar  Rustamov and Johanna  Stachel. 
N.S. acknowledges the support of SERB Ramanujan Fellowship (D.O. No. SB/S2/RJN-084/2015) of the Department of Science and Technology, Government of India.

\newpage
\bibliography{cprs}

\end{document}